\documentclass[twocolumn]{aastex62}

\graphicspath{{./}{figures/}}

\received{XX YY, ZZZZ}
\revised{XX YY, ZZZZ}
\accepted{\today}

\submitjournal{ApJ}

\shorttitle{Properties of NUV-r-J Selected Quiescent Galaxies}
\shortauthors{Hwang et al.}

\begin{document}

\title{Revisiting the Color-Color Selection: Submillimeter and AGN Properties of NUV-r-J Selected Quiescent Galaxies}

\correspondingauthor{Wei-Hao Wang}
\email{whwang@asiaa.sinica.edu.tw}

\author{Yu-Hsuan Hwang}
\affiliation{Physics Department, National Taiwan University, No.\ 1, Sec.\ 4, Roosevelt Rd., Taipei 10617, Taiwan}
\affiliation{Academia Sinica Institute of Astronomy and Astrophysics, No.\ 1, Sec.\ 4, Roosevelt Rd., Taipei 10617, Taiwan}

\author{Wei-Hao Wang}
\affiliation{Academia Sinica Institute of Astronomy and Astrophysics, No.\ 1, Sec.\ 4, Roosevelt Rd., Taipei 10617, Taiwan}

\author{Yu-Yen Chang}
\affiliation{Academia Sinica Institute of Astronomy and Astrophysics, No.\ 1, Sec.\ 4, Roosevelt Rd., Taipei 10617, Taiwan}
\affiliation{Department of Physics, National Chung Hsing University, 40227, Taichung, Taiwan}

\author{Chen-Fatt Lim}
\affiliation{Academia Sinica Institute of Astronomy and Astrophysics, No.\ 1, Sec.\ 4, Roosevelt Rd., Taipei 10617, Taiwan}
\affiliation{Physics Department, National Taiwan University, No.\ 1, Sec.\ 4, Roosevelt Rd., Taipei 10617, Taiwan}

\author{Chian-Chou Chen}
\affiliation{Academia Sinica Institute of Astronomy and Astrophysics, No.\ 1, Sec.\ 4, Roosevelt Rd., Taipei 10617, Taiwan}

\author{Zhen-Kai Gao}
\affiliation{Graduate Institute of Astronomy, National Central University, Taiwan}
\affiliation{Academia Sinica Institute of Astronomy and Astrophysics, No.\ 1, Sec.\ 4, Roosevelt Rd., Taipei 10617, Taiwan}

\author{James S. Dunlop}
\affiliation{Institute for Astronomy, University of Edinburgh, Royal Observatory, Edinburgh EH9 3HJ, UK}

\author{Yu Gao}
\affiliation{Department of Astronomy, Xiamen University, 422 Siming South Road, Xiamen 361005, People’s Republic of China}
\affiliation{Purple Mountain Observatory/Key Laboratory for Radio Astronomy, Chinese Academy of Sciences, 10 Yuanhua Road, Nanjing 210023, People’s Republic of China}

\author{Luis C. Ho}
\affiliation{Kavli Institute for Astronomy and Astrophysics, Peking University, Beijing 100871, Peoples Republic of China}
\affiliation{Department of Astronomy, School of Physics, Peking University, Beijing 100871, Peoples Republic of China}

\author{Ho Seong Hwang}
\affiliation{Korea Astronomy and Space Science Institute, 776 Daedeokdae-ro, Yuseong-gu, Daejeon 34055, Republic of Korea}
\affiliation{Astronomy Program, Department of Physics and Astronomy, Seoul National University, 1 Gwanak-ro, Gwanak-gu, Seoul 08826, Republic of Korea}

\author{Maciej Koprowski}
\affiliation{Institute of Astronomy, Faculty of Physics, Astronomy and Informatics, Nicolaus Copernicus University, Grudziadzka 5, 87-100 Torun,
Poland}

\author{Micha\l{} J. Micha\l{}owski}
\affiliation{Astronomical Observatory Institute, Faculty of Physics, Adam Mickiewicz University, 60-286 Pozna\'{n}, Poland}

\author{Ying-jie Peng}
\affiliation{Kavli Institute for Astronomy and Astrophysics, Peking University, Beijing 100871, Peoples Republic of China}

\author{Hyunjin Shim}
\affiliation{Department of Earth Science Education, Kyungpook National University, Deagu 702-701, Republic of Korea}

\author{James M.\ Simpson}
\affiliation{Centre for Extragalactic Astronomy, Department of Physics, Durham University, UK}
\affiliation{Academia Sinica Institute of Astronomy and Astrophysics, No.\ 1, Sec.\ 4, Roosevelt Rd., Taipei 10617, Taiwan}
\affiliation{National Astronomical Observatory of Japan, Japan}

\author{Yoshiki Toba}
\affiliation{Department of Astronomy, Kyoto University, Kitashirakawa-Oiwake-cho, Sakyo-ku, Kyoto 606-8502, Japan}
\affiliation{Academia Sinica Institute of Astronomy and Astrophysics, No.\ 1, Sec.\ 4, Roosevelt Rd., Taipei 10617, Taiwan}
\affiliation{Research Center for Space and Cosmic Evolution, Ehime University, 2-5 Bunkyo-cho, Matsuyama, Ehime 790-8577, Japan}

\begin{abstract}

We examine the robustness of the color-color selection of quiescent galaxies (QGs) against contamination of dusty star-forming galaxies using latest submillimeter data. We selected 18,304 QG candidates out to $z\sim$ 3 using the commonly adopted $NUV$--$r$--$J$ selection based on the high-quality multi-wavelength COSMOS2015 catalog. Using extremely deep 450 and 850 $\mu$m catalogs from the latest JCMT SCUBA-2 Large Programs, S2COSMOS and STUDIES, as well as ALMA submillimeter, VLA 3 GHz, and \emph{Spitzer} MIPS 24 $\mu$m catalogs, we identified luminous dusty star-forming galaxies among the QG candidates. We also conducted stacking analyses in the SCUBA-2 450 and 850 $\mu$m images to look for less-luminous dusty galaxies among the QG candidates. By cross-matching to the 24 $\mu$m and 3 GHz data, we were able to identify a sub-group of ``IR-radio-bright'' QGs who possess a strong 450 and 850 $\mu$m stacking signal. The potential contamination of these luminous and less-luminous dusty galaxies account for approximately 10\% of the color-selected QG candidates. In addition, there exists a spatial correlation between the luminous star-forming galaxies and the QGs at a $\lesssim60$ kpc scale.  Finally, we found a high QG fraction among radio AGNs at $z<$ 1.5. Our data show a strong correlation between QGs and radio AGNs, which may suggest a connection between the quenching process and the radio-mode AGN feedback.

\end{abstract}


\section{Introduction} \label{sec:intro}

Quiescent galaxies (QGs) are defined as galaxies with their star formation rates (SFR) lower than the average SFR of star-forming galaxies (SFGs) of similar stellar masses at similar redshifts \citep[i.e., below the ``main sequence'';][]{Daddi2007,Elbaz2007,Noeske2007,Wuyts2011,Ciambur2013}. The quenching mechanism of QGs is an important topic, especially for high-redshift ones. Observations showed that populations of massive QGs exist at $z \sim$ 1.0--2.0 \citep{Belli2017, Carnall2019, Newman2018}, and that half of the most massive QGs were formed at $z \sim$ 1.5 \citep{Ilbert2013, Muzzin2013}. In recent studies, the QG population has been extended to $z \sim$ 4.0 \citep{Straatman2014, Glazebrook2017, Merlin2018, Schreiber2018a, Carnall2020, Forrest2020a, Forrest2020b, Valentino2020}. Among these, \citet{Valentino2020} reported three QGs with stellar masses around $10^{11}$ M$_{\odot}$ and with SFR ranging from 1.1 to 24.0 $M_{\odot}$ year$^{-1}$ (1.0 to 2.1 $\sigma$ below the main sequence) at $z$ = 3.775, 4.012, and 3.767.  How these high-$z$ QGs can increase their mass and quench the star formation in a short period is still not fully understood. 

Active galactic nucleus (AGN) feedback is one of the proposed scenarios to explain the rapid quenching \citep{Bower2006, Croton2006, Somerville2008, Fabian2012, Man2018}. AGN activities may either provide kinetic energy to the interstellar medium in the host galaxies and reduce the star formation efficiency, or heat up the gas to prevent the gas from cooling \citep[radio mode AGN feedback;][]{Croton2006, Bower2006, Fabian2012, Somerville2008}. AGN activities may also remove the gas and terminate the star formation \citep[quasar mode AGN feedback;][]{Fabian2012, Somerville2008}.

However, there are also other theoretical scenarios for the quenching mechanism. For example, the existence of turbulence may also provide kinetic energy to the system and result in morphological quenching \citep{Martig2009,Dekel2009}. Another example would be positive AGN feedback. In this case, AGN activities enhance star formation but rapidly consume the gas, resulting in a quenched galaxy \citep{Ishibashi2012, Zhuang2020, Shangguan2020}. Other theoretical scenarios include mergers, stellar feedback, virial shock heating, etc \citep[see discussion in][and references therein]{Man2018}, and the quenching of massive galaxies could be a combination of some of these scenarios. Which of the mentioned scenarios is the dominant channel of the quenching mechanism remains unclear. 

Here we would like to first focus on the foundation of the QG studies, the selection criteria of QGs. Rest-frame color-color diagrams are widely applied to selecting QGs. They are often composed of two rest-frame colors: a UV-to-optical color (typically as the $y$-axis) to distinguish blue SFGs from red QGs using the strong UV emission from young stars, and an optical-to-near-infrared color (typically as the $x$-axis) to distinguish old passive stellar populations from dusty/reddened young stellar populations. Such photometric selections are very convenient since only photometric data are needed. There are various kinds of color-color diagrams proposed for QG selections based on rest-frame absolute magnitudes, such as the $U$--$V$--$J$ diagram \citep{Wuyts2007,Williams2009,Muzzin2013}, the $NUV$--$r$--$J$ diagram \citep{Ilbert2013}, and the $NUV$--$r$--$K$ diagram \citep{Arnouts2013}. 

Despite the great success of using color-color diagrams to select large samples of QGs, there are potential issues in such selections. First, the selection boundary that separates QGs and SFGs in a color color diagram was decided empirically \citep[e.g.,][]{Ilbert2013, Muzzin2013, Williams2009}. Whether a set of selection criteria are still applicable to different datasets should be examined. Second, either because of the intrinsic properties of SFGs and QGs, or because of photometric errors, the distribution of the two groups of galaxies can have unknown levels of overlap around the selection boundaries. It was found that adjusting the position of the boundary by $\pm$0.1 mag could greatly change the selection efficiency and the subsequent analyses based on the selected samples \citep[][Appendix B therein]{Muzzin2013}.

One important factor here is that our understanding of dusty galaxies is limited. The spectral energy distribution (SED) of these high-$z$ dusty galaxies may be more complicated than what was initially assumed to setup color selection.  The selected QG ``candidates'' may still suffer from contamination by red dusty SFGs at high $z$. For instance, a $z$ = 3.717 QG candidate \citep{Straatman2014,Glazebrook2017} was detected at 450 and 870 $\mu$m \citep{Simpson2017}. This implies that the target is a dusty SFG, or is an interacting system consisting of a QG and a dusty galaxy \citep[e.g.,][]{Schreiber2018a}, or at least contains a significant dusty star-forming component. Such contamination may lead to an overestimated number density of QGs, especially at high $z$, which may have consequences in our pursuit of an understanding of the quenching mechanism.

Therefore, in this paper, we would like to revisit the selection of QGs using color-color diagrams. We will examine the quiescence of our color-selected QG candidates using submillimeter observations. Various previous studies were carried out to analyze the star formation of color-selected QGs. Some studies measured the SFR by either applying SED fittings from the UV to mid-infrared bands \citep{Fumagalli2014, Merlin2018, Carnall2019, Toba2019}, using spectroscopic data \citep{Schreiber2018,Belli2017}, or measuring H$\alpha$ emission \citep{Belli2017b}. Others searched for far-infrared dust emissions using $Spitzer$ observations \citep{Fumagalli2014, Man2016, Gobat2018, Magdis2021}, $Herschel$ observations \citep{Viero2013, Man2016,Straatman2014, Merlin2018, Gobat2018, Magdis2021} or Atacama Large Millimeter/submillimeter Array (ALMA) observations \citep{Santini2019, Schreiber2018, Simpson2017}. Others also searched for gas content in QG candidates \citep{Sargent2015,Young2011}. In particular, \citet{Man2016} measured the dust emission of their color-selected QGs by stacking $Herschel$ SPIRE data at 250, 350, and 500 $\mu$m (FWHM $\simeq$ $18\arcsec$.2, $24\arcsec$.9, and $36\arcsec$.3), and they claimed that contamination of dusty SFG is $\sim$ 15 \% among their QG candidates. We will follow the approach in \citet{Man2016} but re-examine the issue with higher angular resolutions using JCMT SCUBA-2 450 and 850 $\mu$m data (FWHM = $7\arcsec$.9 and $13\arcsec$) and ALMA data.

In this study, we selected 18,304 QG candidates using the $NUV$--$r$--$J$ diagram with deep galaxy samples from the COSMOS field \citep{Laigle2016} and analyzed the properties of the selected QG candidates. In the first part of our study, we estimated the contamination of dusty SFGs among the QG candidates by cross-matching them to the multi-wavelength catalogs as well as performing stacking analyses in the submillimeter images. We also estimated the effect of chance projection in the cross-matching and estimated the degrees of small-scale clustering. In the second part of our study, we further investigated the AGN feedback as a potential quenching mechanism among QG candidates. We examined the relation between various AGNs and the QG candidates in our data by calculating the QG fractions in the AGN samples.

We describe our data in Section~\ref{sec:data} and introduce the QG color-color selection in Section~\ref{sec:QG_selection}. In Sections~\ref{sec:bright_SMG} and \ref{sec:faint_SMG}, we analyze the contamination of dusty SFGs among the $NUV$--$r$--$J$ selected QG candidates. We also examine the small-scale clustering between dusty SFGs and QG candidates in Section~\ref{subsec:blind_matching}. In Section~\ref{sec:AGN_properites}, we discuss the QG fractions among our AGN samples. Section~\ref{sec:summary} gives a summary of our results. We use the \citet{Chabrier2003} initial mass function (IMF) and an H$_0$ = 70 km s$^{-1}$ Mpc$^{-1}$, $\Omega_\Lambda$ = 0.7, and $\Omega_m$ = 0.3 cosmology throughout this study.

\section{Multi-wavelength Data} \label{sec:data}

\subsection{COSMOS2015 Catalog} \label{subsec:COSMOS2015}

We selected galaxies from the multi-wavelength band-merged COSMOS2015 catalog \citep{Laigle2016}. To use the rest-frame absolute magnitudes for our color selection, we excluded those labeled as failure in SED fitting to avoid bad absolute magnitudes.  We also excluded samples that are labeled as stars.  We excluded galaxies with extreme values in the catalog ($NUV$, $r$, and $J$ absolute magnitudes that are $<-30$ or $>0$, and negative redshifts), which are likely caused by either catastrophic failures in SED fitting or problematic photometry. These selection criteria reject $\sim57\%$ (677,085/1,182,108) of the initial sample.

 We further limited the errors of the magnitudes in $K_S$ band to be lower than 0.2. This uniform selection ensures that our sample has a robust set of photometry and avoids biasing against high-$z$ sample. The limiting magnitude is 23.7 for $K_S$ band. The selection criterion of the $K_S$ band magnitude error further rejects $\sim29\%$ (344,806/1,182,108) of the initial sample. Overall, the majority of the rejections are caused by their faintness. They either are not detected at $K_S$ or have $K_S>24$.

\begin{figure}[ht!]
\epsscale{1.15}
\plotone{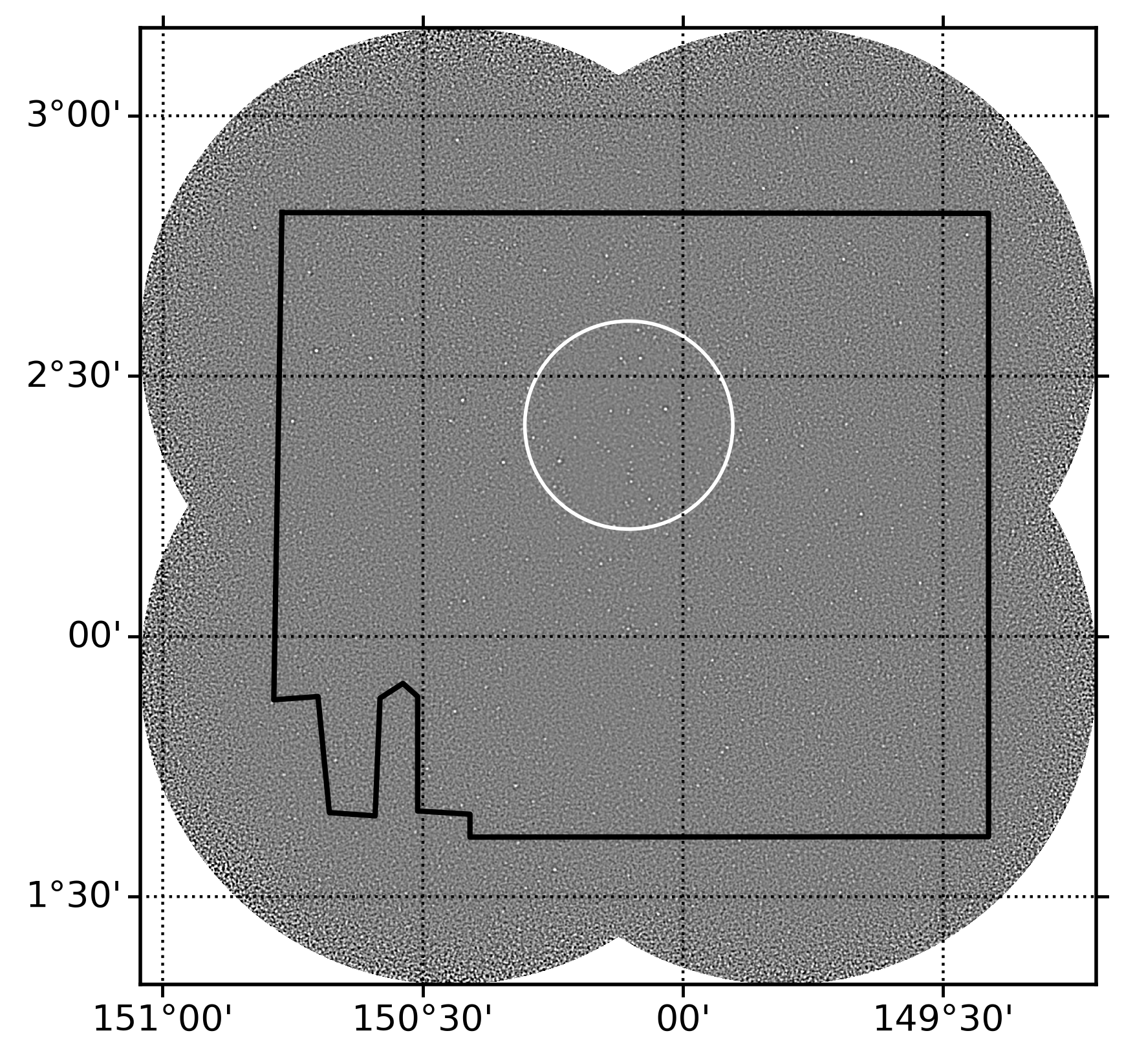}
\caption{\label{fig:footprint}Coverage maps of the COSMOS field. The background shows the S2COSMOS 850 $\mu$m image. The black polygon corresponds to the coverage of our $K_s$ band selected COSMOS2015 sample, while the white circle corresponds to the 151-arcmin$^2$ coverage of STUDIES 450 $\mu$m image. The MIPS 24 $\mu$m and VLA 3 GHz catalogs cover the whole area of the black polygon and are not shown in this figure.}
\end{figure}

\begin{figure*}[ht!]
\epsscale{1.15}
\plottwo{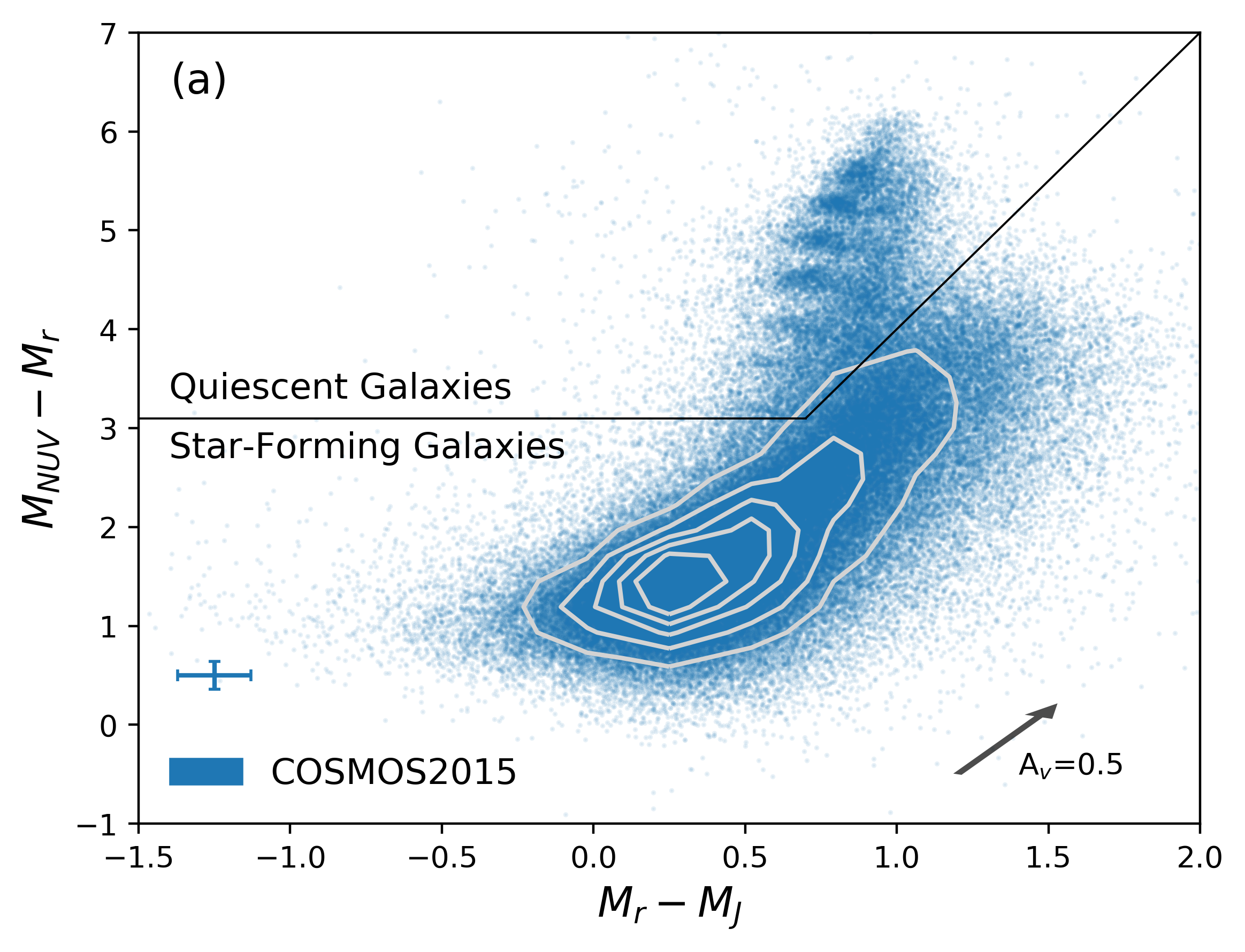} {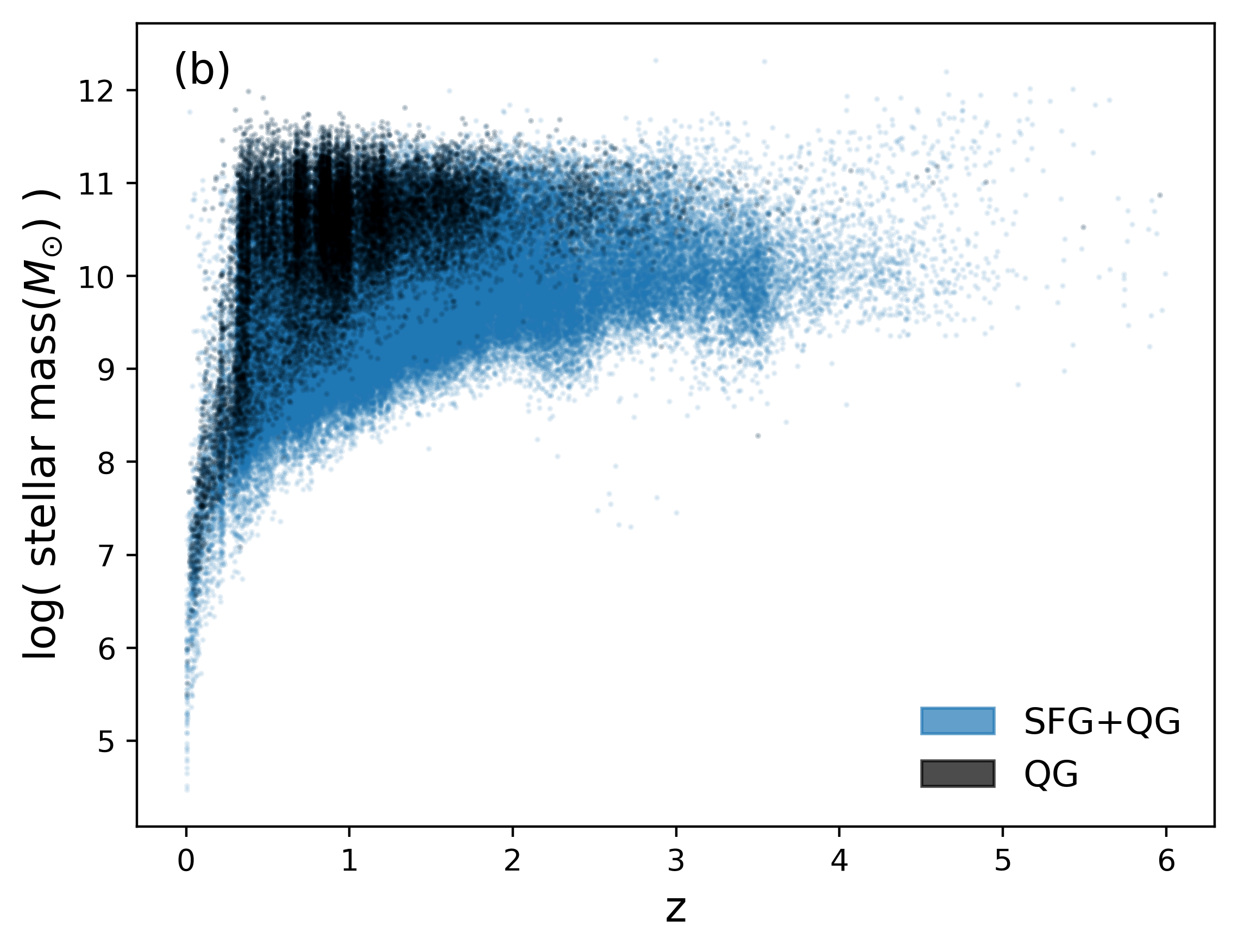}
\plottwo{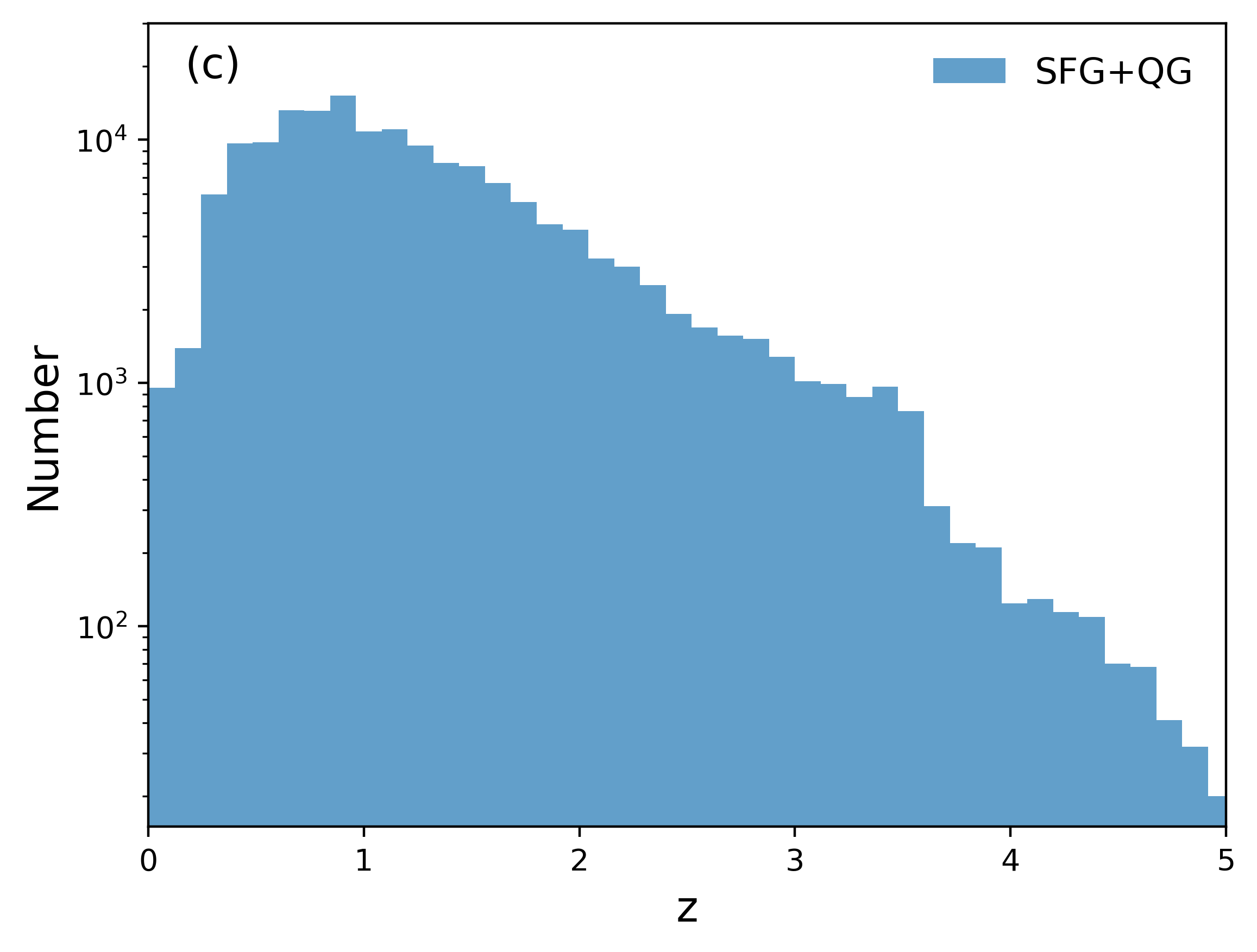}{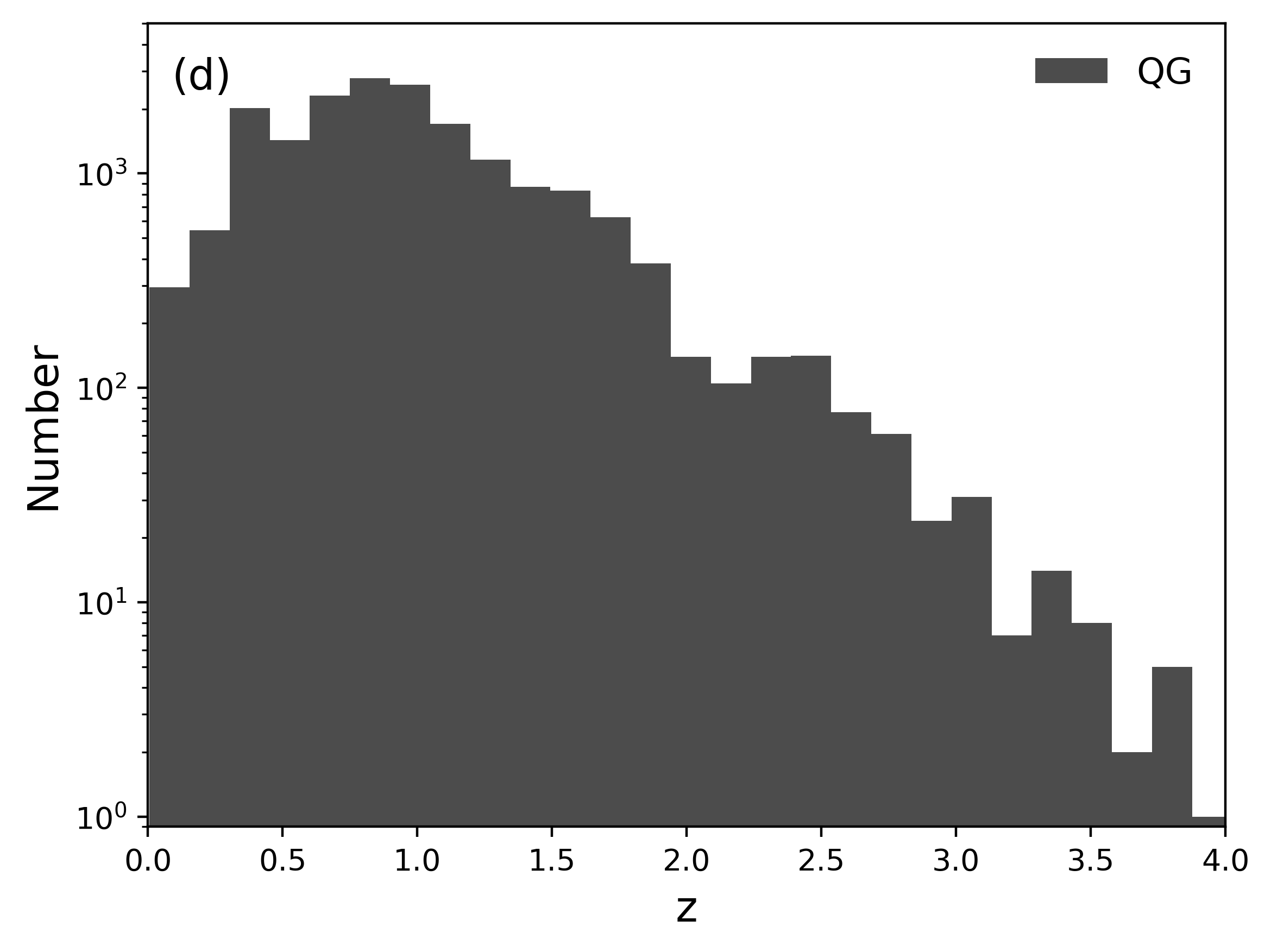} 
\caption{\label{fig:data}The full galaxy sample selected from the COSMOS2015 catalog. Panel (a) shows the distribution in the $NUV$--$r$--$J$ diagram, and the QG sub-sample is selected in upper-left corner. The fringe structure in the QG population may be caused by either the lack of QG template or by certain procedures in the SED fitting, but the structure does not affect our result. The reddening vector derived from \citet{Calzetti2000} extinction is shown in the lower-right corner, while the typical (median) error in the two colors for all the sources is shown in the lower-left corner. Panel (b) and (c) show the stellar mass and the redshift distributions of the full galaxy sample (colored in blue), while panel (b) and (d) show those of the QG sub-sample (colored in black).}
\end{figure*}

With the above selection criteria, we obtained a total sample size of 160,217 galaxies from the COSMOS2015 catalog. They all have high-quality SED fitting results; all of them have SED fitting based on at least nine filters, and 98\% of them more than 28 filters. The sample covers an area of 1.58 deg$^2$ in the COSMOS field (Fig.~\ref{fig:footprint}). We used stellar mass $M_*$, photometric redshift $z$, and rest-frame absolute magnitudes $M_{NUV}$, $M_r$, and $M_J$ from the COSMOS2015 catalog, which were derived from SED fittings. The sample has stellar masses up to $M_* = 10^{12}~M_{\odot}$ and redshifts over $z\sim4$ (Fig.~\ref{fig:data} (b) and (c)). The absolute magnitudes will be applied for our QG selection in the next section.

\subsection{Submillimeter Data} \label{subsec:submm}

We used submillimeter data in the COSMOS field from JCMT SCUBA-2 \citep{Holland2013, Holland1999} at 450 $\mu$m (STUDIES, \citealt{Wang2017}; final data release in Gao et al.\ 2021, in prep.) and 850 $\mu$m (S2COSMOS, \citealt{Simpson2019}), in order to search for dusty SFGs that contaminate the QG sample. The 450 $\mu$m map covers the central 151 arcmin$^2$ of COSMOS, while the 850 $\mu$m map covers the whole COSMOS field (Fig.~\ref{fig:footprint}). The 450 $\mu$m and 850 $\mu$m maps have detection limits of about 3.5 mJy and 2 mJy, respectively. The detection limits are all substantially higher than the confusion noise ($\sigma_{\rm c}\sim0.7$ mJy at 450 $\mu$m, e.g., \citealp{Lim2020}; $\sigma_{\rm c}\sim0.5$ mJy at 850 $\mu$m, e.g., \citealt{Simpson2019}).

In total, we selected 357 objects with 450 $\mu$m detection and 1,147 objects with 850 $\mu$m detection from the SCUBA-2 maps. Four of the 450 $\mu$m sources and 166 of the 850 $\mu$m sources located outside the region occupied by our optically selected sample, because of the difference in area coverage and the masks in the COSMOS2015 catalog (Fig.~\ref{fig:footprint}). Among the remaining 353 objects with 450 $\mu$m detection and 981 objects with 850 $\mu$m detection, 77 and 370, respectively, have ALMA observations from the AS2COSMOS and A3COSMOS catalogs (Section~\ref{subsec:auxiliary}).

Since the SCUBA-2 maps have relatively low angular resolution, we could not reliably identify the optical counterparts to the submillimeter sources. We therefore include the auxiliary data in section \ref{subsec:auxiliary} for the process of cross matching COSMOS2015 galaxies to SCUBA-2 sources, and the details will be described in section \ref{subsec:traditional_matching}.

\subsection{Auxiliary Data} \label{subsec:auxiliary}

We included $Spitzer$ MIPS 24 $\mu$m, VLA 3 GHz, and ALMA catalogs in our study for their better astrometry when our submillimeter data do not have sufficient angular resolution and for analyzing QG properties.

For 24 $\mu$m data, we used the $Spitzer$ MIPS S-COSMOS image from \citet{Sanders2007}. In order to generate a catalog deeper than the archival MIPS catalog of \citet{Sanders2007}, 24 $\mu$m sources were extracted  using \texttt{SExtractor} \citep{Bertin1996}, and their fluxes were re-calibrated to their $Spitzer$ General Observer Cycle 3 total fluxes. Our MIPS 24 $\mu$m catalog has a 3.5$\sigma$ detection limit of  57 $\mu$Jy, in contrast to the flux cut at 150 $\mu$Jy in \citet{Sanders2007}. Our catalog is very similar to the catalog of \citet{LeFloch2009} in terms of total numbers of detections. The fluxes are also consistent within 6\% \citep{Lim2020} as we calibrated our fluxes to that of  \citet{Sanders2007}.

We cross-matched the COSMOS2015 catalog with our MIPS 24 $\mu$m catalog using a search radius of $2\arcsec$, which corresponds to about 1/3 of the beam size at 24 $\mu$m. 26,999 galaxies (16.9\%) are matched to the MIPS 24 $\mu$m sources.

For 3 GHz data, we directly adopted the identification of the COSMOS2015 objects in the VLA catalog of \citet{Smolcic2017a}, which used a search radius of $0\farcs8$. The 5$\sigma$ detection limit of the VLA catalog of \citet{Smolcic2017a} is 2.3 $\mu$Jy beam$^{-1}$. 6,002 galaxies (3.7\%) are matched to the VLA 3-GHz sources.

We also used catalogs derived from ALMA observations, including the AS2COSMOS \citep{Simpson2020} and A3COSMOS \citep{Liu2019} catalogs. The AS2COSMOS catalog was derived from the follow-up 343 GHz observations of 186 bright 850 $\mu$m sources in the S2COSMOS catalog. The AS2COSMOS sources are essentially complete for the S2COSMOS sources above 6.2 mJy; only one S2COSMOS source does not have ALMA detection. The A3COSMOS catalog collects ALMA archival data in the COSMOS field, at wavelengths from 671 to 90.2 GHz. We cross-matched our optical sample with the ALMA catalogs using a search radius of $1\arcsec$. This search radius should allow us to overcome the intrinsic offsets between starlight and submillimeter emission from dusty galaxies (e.g., 1$\sigma$ offset of $0\farcs55$ in \citealp{Chen2015}).

\subsection{AGN Sample} \label{subsec:AGN_samples}

We also examined the AGN properties of our sample. We cross-matched our sample with radio AGNs from the VLA catalog of \citet{Smolcic2017}, color-selected mid-IR AGNs from \citet{Chang2017}, and X-ray AGNs selected from $Chandra$ data by \citet{Civano2016} and \citet{Marchesi2016}.

The radio AGNs were selected by comparing the observed radio emission to the expected radio emission from IR-derived SFR. Those exceeding 3$\sigma$ in $log(L_{1.4GHz}/\mathrm{SFR}_{IR})$ are classified as radio AGNs \citep[see the details in][]{Smolcic2017, Delvecchio2017}. The mid-IR AGNs were selected in the rest-frame mid-IR color-color diagram. Those which exhibit red power-law SEDs in the mid-IR are classified as mid-IR AGNs \citep[see the details in][]{Chang2017,Lacy2004, Lacy2007, Donley2012}. The X-ray AGNs were selected with X-ray luminosity of $L_{X(2-10keV)}>10^{42}$ ergs s$^{-1}$ \citep{Zezas1998,Ranalli2003,Szokoly2004}. We note that if such an X-ray luminosity is produced purely by X-ray binaries rather than an AGN, the inferred SFR would be $>200$~$M_{\odot}$~yr$^{-1}$ using the conversion between SFR and $L_{X}$ \citep[e.g.,][]{Ranalli2003}. Such a high SFR would be detect in our submillimeter analyses, but we do not observe it. Therefore, the majority of the $L_{X(2-10keV)}>10^{42}$ ergs s$^{-1}$ sources in our samples should be AGN-dominated.

\section{Color-Color Diagram} \label{sec:QG_selection}

We applied the rest-frame $NUV$--$r$--$J$ color-color diagram to our sample in order to select QG candidates. Various color-color diagrams were used for QG selection, including the $U$--$V$--$J$ diagram \citep{Williams2009} and the $NUV$--$r$--$J$ diagram \citep{Ilbert2013}. Although the $U$--$V$--$J$ diagram is more widely used than the $NUV$--$r$--$J$ diagram, there are advantages of using $NUV$ and $r$ bands instead of $U$ and $V$ bands \citep{Ilbert2013}. The $NUV$ band is at a shorter wavelength, so it is more sensitive to emission from young stars and extinction. The $NUV-r$ color has a wider wavelength span than the $U-V$ color, so it is less vulnerable to photometric errors. The rest-frame $NUV$ band can be obtained from optical data toward higher redshifts, around $z >$ 2, where $U$ band starts to enter the near-IR.  This leads to better sensitivities. Because of the above, we adopted the $NUV$--$r$--$J$ diagram in this study. We note that the selection results of using the two color-color diagrams are similar to each other. About 85\% of our $NUV$--$r$--$J$ selected QG candidates overlap with the $U$--$V$--$J$ selected QG candidates, and the overlapping fraction slightly varies with redshift and the position of the selection boundary.

On the $NUV$--$r$--$J$ color-color diagram (Fig.~\ref{fig:data} (a)), a blue color in $y$-axis indicates the starlight from young stars, while the color in $x$-axis breaks the degeneracy between age and dust reddening. QGs tend to locate in the upper-left corner of the diagram. We adopted the criteria proposed by \citet{Ilbert2013}:
$$M_{NUV}-M_{r}> 3(M_{r}-M_{J})+1,$$ 
$$M_{NUV}-M_{r}> 3.1.$$

\begin{figure}[ht!]
\epsscale{1.05}
\plotone{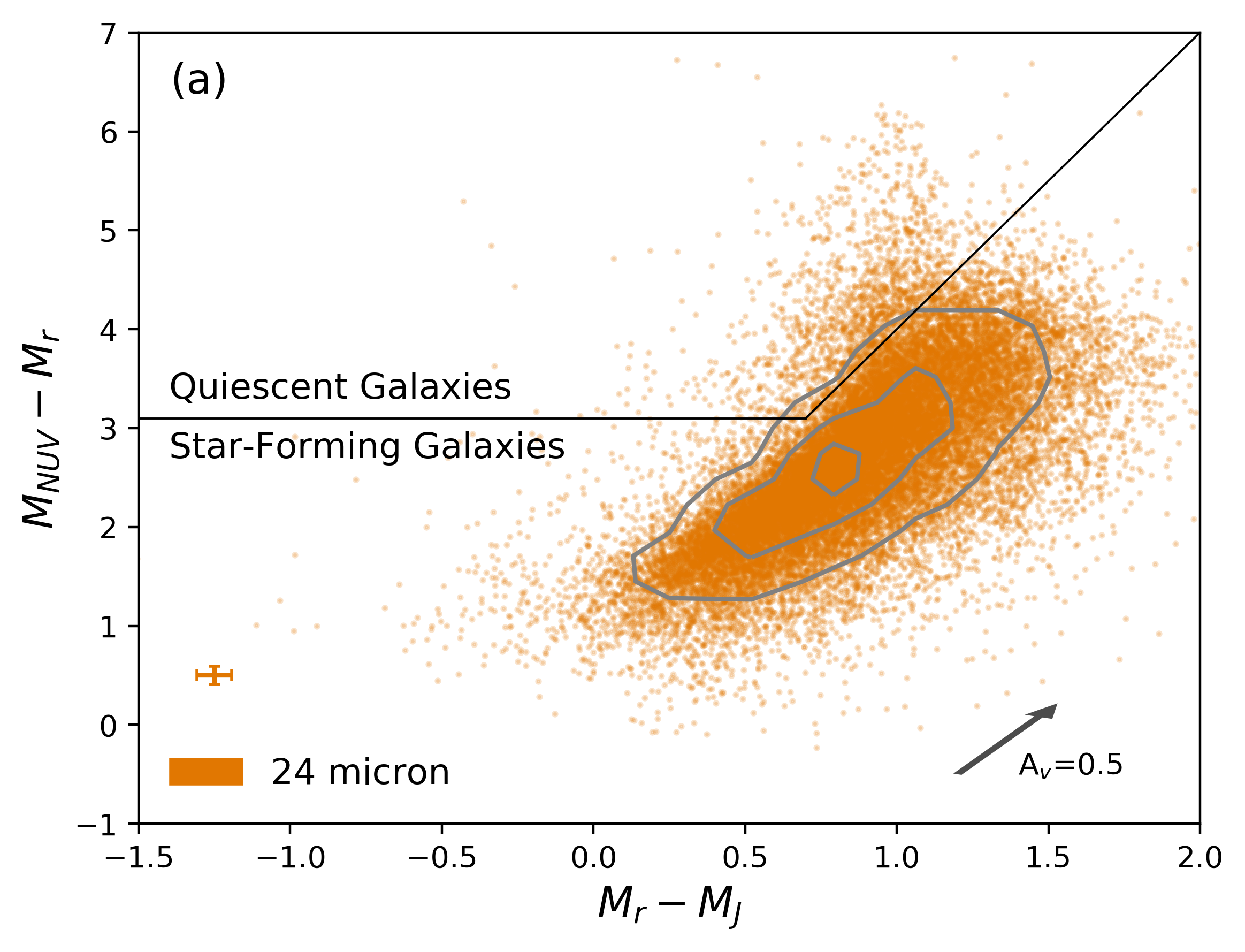}
\epsscale{1.05}
\plotone{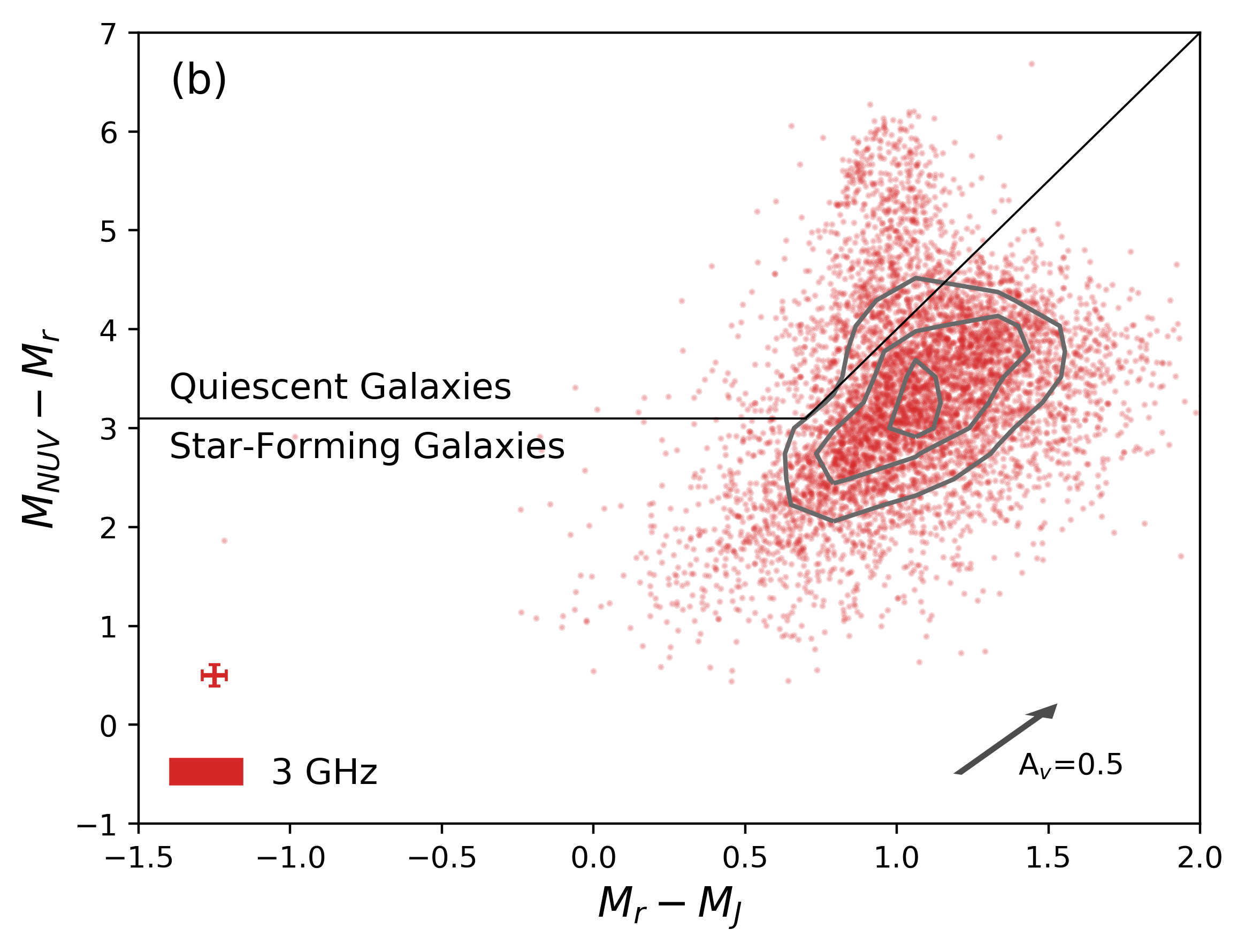}
\epsscale{1.05}
\plotone{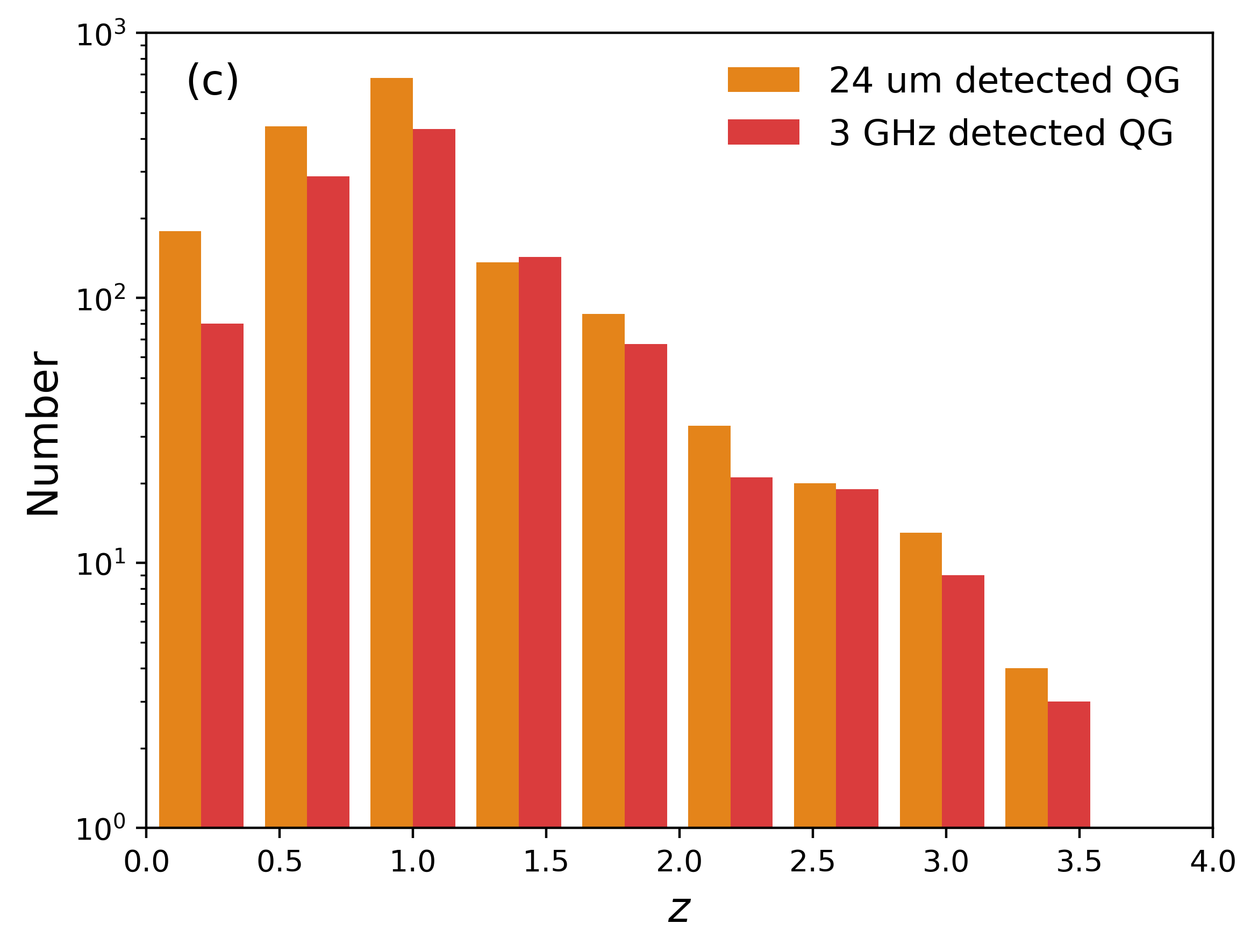}
\caption{\label{fig:NUVrJ243}Distribution of the 24-$\mu$m detected sample (a) and the 3-GHz detected sample (b) in the $NUV$--$r$--$J$ diagram. The reddening vector derived from \citet{Calzetti2000} extinction and the typical errors in the two colors are also shown, as in Fig.~\ref{fig:data} (a). The typical errors are smaller than that of the full galaxy sample (Fig.~\ref{fig:data}(a)). This is resulted from the higher fractions of bright galaxies among the two subgroups (median $r$ $\sim0.5$ magnitudes brighter).} The QG candidates are selected in upper-left corner of the panels, and the redshift distribution of the two QG subgroups are shown in (c). We note that the two subgroups have partial overlap between each other.
\end{figure}

We selected 18,304 galaxies to be our QG candidates, which are 11.4$\pm$0.1 \% of the total (Fig.~\ref{fig:data}(a)). The selected QG candidates have a redshift distribution peaking at $z\sim1.0$ and extending to $\sim3.0$ (Fig. \ref{fig:data} (d)). Our selection result is well consistent with the flag ``CLASS=0'' in the COSMOS2015 catalog \citep{Laigle2016}, which applied the same $NUV$--$r$--$J$ selection method. Among the 24 $\mu$m detected galaxies, 5.9$\pm$0.1 \% enter the QG selection region and therefore are QG candidates (Fig. \ref{fig:NUVrJ243} (a)). Among the 3 GHz detected galaxies, 17.8$\pm$0.5 \% are QG candidates (Fig. \ref{fig:NUVrJ243} (b)). The redshift distributions of the 24 $\mu$m and 3 GHz detected QGs also peak at $z\sim1.0$ but have larger fraction of QGs at high $z$  (Fig. \ref{fig:NUVrJ243} (c)). The QG selection of submillimeter-detected galaxies will be described in Section \ref{subsec:traditional_matching}. The numbers of selected QG candidates are summarized in Table \ref{tab:data}.

To better understand the diagrams, we show the reddening vector and the typical (median) errors of the two colors in Fig.~\ref{fig:data} (a), Fig.~\ref{fig:NUVrJ243} (a), and Fig.~\ref{fig:NUVrJ243} (b). The reddening vector is derived from \citet{Calzetti2000} extinction. For the magnitude errors, unfortunately the COSMOS2015 catalog does not provide errors in the absolute magnitudes. To have a rough idea of the errors, we followed the COSMOS2015 procedure (O.\ Ilbert \& I.\ Davidzon, private communication) to select the nearest broad-band filter in the rest-frame that has a photometric error of $<0.3$. And we use the photometric error of that filter band as the absolute magnitude error. This clearly does not account for the errors in the $K$-corrections derived from the fitted SEDs, nor the errors propagated from the photo-$z$ errors, but should still include a substantial part of the error budget.

With the above-estimated photometric errors, we could further estimate the fraction among the QG candidates that may originate from the SFG color space and scattered into the QG color space by the photometric errors. For each QG candidate, we generated 1,000 randomly perturbed $NUV$, $r$, and $J$ band absolute magnitudes that follow Gaussian distributions according to the photometric errors. We then calculated the percentage of the perturbed colors that are located in the SFG region, i.e., the probability of the QG candidate to be selected as a SFG if there were no photometric errors. The average probability among our QG candidates is $\sim$7.5\%, meaning that $\sim$7.5\% of our selected QG may have moved from the SFG color space across the boundary into the QG color space due to their photometric errors.  The probabilities can help to understand the nature of dusty SFG contamination in the color-selected QG population. We will further discuss this in Sections \ref{sec:bright_SMG} and \ref{sec:faint_SMG}.

\begin{deluxetable*}{l|llccc}
\tablecaption{\label{tab:data}Sample sizes and results of multi-wavelength cross-matching.}
\tablehead{
\colhead{} & \colhead{SFGs+QGs} & \colhead{QGs} & \colhead{QGs/(all SFGs+QGs)} & \colhead{QGs/(all QGs)} & \colhead{QGs by chance projection}
}
\startdata
total in the COSMOS field & 160217 & 18304 & 11.4$\pm$0.1 \% & 100 \% & - \\
24 um detected & 26999 & 1596 & 5.9$\pm$0.1 \% & 8.72$\pm$0.22 \% & 382 \\
3 GHz detected & 6002 & 1066 & 17.8$\pm$0.5 \% & 5.82$\pm$0.18 \% & - \\
850 $\mu$m detected & 653 & 30 & 4.6$\pm$0.8 \% & 0.16$\pm$0.03 \% & 7.0\\
~~~850 $\mu$m + ALMA & 289 & 11 & 3.8$\pm$1.1 \% & - & 1.3\\
~~~850 $\mu$m + 24 $\mu$m + 3 GHz & 364 & 19 & 5.2$\pm$1.2 \% & - & 5.8\\
\hline
total in the STUDIES field & 15296 & 1846 & 12.1$\pm$0.3 \% & 100 \% & - \\
450 $\mu$m detected & 239 & 8 & 3.3$\pm$1.2 \% & 0.43$\pm$0.15 \% & 2.5\\
~~~450 $\mu$m + ALMA & 58 & 2 & 3.4$\pm$2.4 \% & - & 0.3\\
~~~450 $\mu$m + 24 $\mu$m + 3 GHz & 181 & 6 & 3.3$\pm$1.4 \% & - & 2.1\\
\hline
radio AGN & 1378 & 563 & 40.9$\pm$1.7 \% & 3.08$\pm$0.13 \% & - \\
mid-IR AGN & 791 & 95 & 12.0$\pm$1.2 \% & 0.52$\pm$0.05 \% & - \\
X-ray AGN & 2267 & 413 & 18.2$\pm$0.9 \% & 2.26$\pm$0.11 \% & - \\
\enddata
\tablecomments{The errors are set to be Poissonian, and only reflect the uncertainties caused by the finite sample sizes. The 850 $\mu$m and 450 $\mu$m detected samples are determined through both the low-resolution SCUBA-2 data and the high resolution auxiliary data. The auxiliary data are either ALMA data, or 24 $\mu$m and 3 GHz data (see Section \ref{subsec:traditional_matching} for details).}
\end{deluxetable*}

From Fig. \ref{fig:NUVrJ243} (a), we can see that most of the 24 $\mu$m detected QG candidates tend to distribute close to the selection boundary in the diagram. They can be either dusty galaxies entering the QG color space because of atypical SED shapes, simply regular SFGs scattered into the QG color space because of photometric errors (cross in Fig. \ref{fig:NUVrJ243} (a)), or chance projections in the cross matching. By measuring the search area of matching through $2\arcsec$ search radius, we estimated that 382$\pm$20 out of the 1596 matches (23.9$\pm$1.2\%) can be chance projections. In Table \ref{tab:data}, 24 $\mu$m detected galaxies have lower QG fractions than that of all the COSMOS2015 sample. 24 $\mu$m sources are sensitive to dust emission, and the low fraction suggests a low dusty-galaxy contamination in the QG color selection.

On the other hand, the 3 GHz detected QG candidates distribute well into the QG selection region in Fig. \ref{fig:NUVrJ243} (b). If we calculate their vertical distances to the selection boundary, we obtain median values of 0.4 and 0.7 for the 24 $\mu$m and 3 GHz detected QG candidates, respectively. The median distance for the 3 GHz detected QGs is much larger than the typical photometric error (cross in Fig. \ref{fig:NUVrJ243} (b)), so they are not SFGs scattered into the QG color space. A large fraction of them should be real QGs harboring radio AGNs (see Section \ref{sec:AGN_properites} and Fig.~\ref{fig:NUVrJ_AGN} for further evidence). In Table \ref{tab:data}, the QG fraction among them is considerably higher than those of all the other subgroups. This gives us a hint about the correlation between radio AGN and QG candidates, which will be discussed in Section \ref{sec:AGN_properites}.

\section{Bright Submillimeter Galaxies Among QG Candidates} \label{sec:bright_SMG}

In this section, we conduct a thorough analysis on the contamination of bright submillimeter galaxies among our QG candidates. In Section \ref{subsec:traditional_matching}, we cross-matched our sample with the SCUBA-2 450 $\mu$m and 850 $\mu$m catalogs using the positions of MIPS 24 $\mu$m, VLA 3 GHz, and ALMA submillimeter sources. In Section \ref{subsec:blind_matching}, we further performed a blind cross-matching and reported the finding of small-scale clustering between QG candidates and SCUBA-2 sources.

\subsection{Traditional Cross Matching} \label{subsec:traditional_matching}

\subsubsection{Counterpart Identification using Auxiliary Data} \label{subsubsec:traditional_matching_process}

We have searched for MIPS 24 $\mu$m, VLA 3 GHz, and ALMA counterparts in the COSMOS2015 catalog with data presented in Section \ref{subsec:auxiliary}. We can therefore search for the optical counterparts to the low-resolution SCUBA-2 submillimeter sources by including the high-resolution multi-wavelength information. Such a two-step counterpart identification method is traditionally used on SCUBA-2 sources. In general, this method was shown to be able to pick up some 2/3 of SCUBA-2 source counterparts \citep[e.g.,][]{Casey2013,Koprowski2016,Cowie2017,Michalowski2017,An2018,Simpson2020,Lim2020}, but the exact fractions depend on the sensitivity of the high-resolution observations in the mid-IR, submillimeter, or radio. 

We first cross-matched our optical sample with the SCUBA-2 450 $\mu$m and 850 $\mu$m sources using a search radius of $4\arcsec$ and $7\arcsec$, respectively. The search radii are approximately half of the full width at half maximum of the beams (FWHM = $7\arcsec$.9 at 450 $\mu$m and $13\arcsec$ at 850 $\mu$m). Such larger search radii (cf.\ 1/3 FWHM for the 24 $\mu$m matching) are required as the SCUBA-2 positional accuracy is more impacted by confusion effects and telescope pointing errors, rather than just the beam sizes. Then, from the matched sample, we narrowed down the optical counterparts by searching for ALMA detected galaxies from the AS2COSMOS and A3COSMOS catalogs (described in Section \ref{subsec:auxiliary}). For the remaining sources without ALMA detection, we identified their optical counterparts by searching for 24 $\mu$m and 3 GHz detected galaxies (described in Section \ref{subsec:auxiliary}). Those without MIPS and VLA counterparts are likely to be at higher redshifts \citep[$z\gtrsim3$, see Section 3.3 in][]{Lim2020} and are not the main targets of interest in this paper given the redshift distributions in Fig.~\ref{fig:data}. We note that when there are multiple sources within the search radius, we consider all of the sources and narrow down the possible counterparts only with multi-wavelength information without considering their distances to the SCUBA-2 position.

\begin{figure}[ht!]
\epsscale{1.15}
\plotone{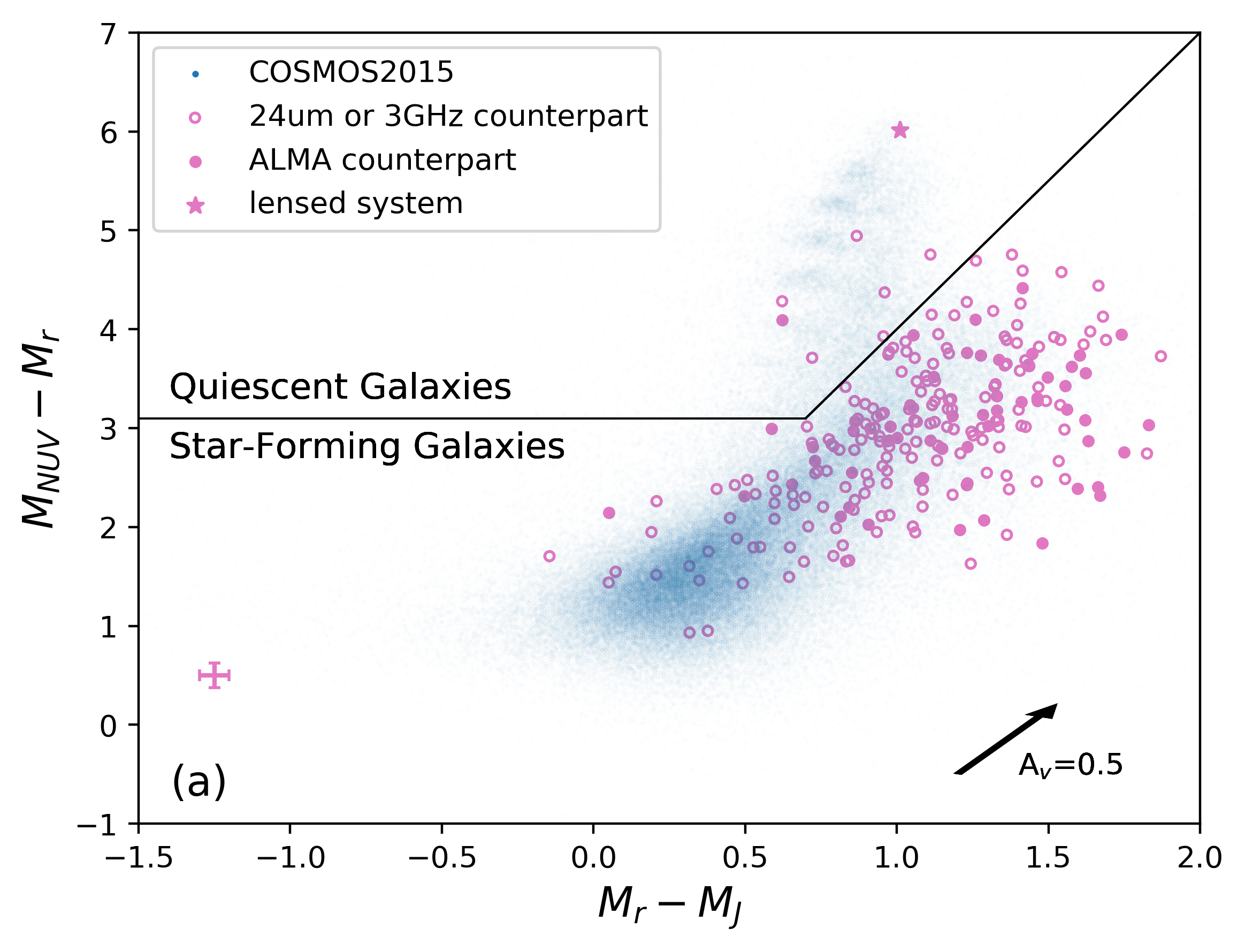}
\plotone{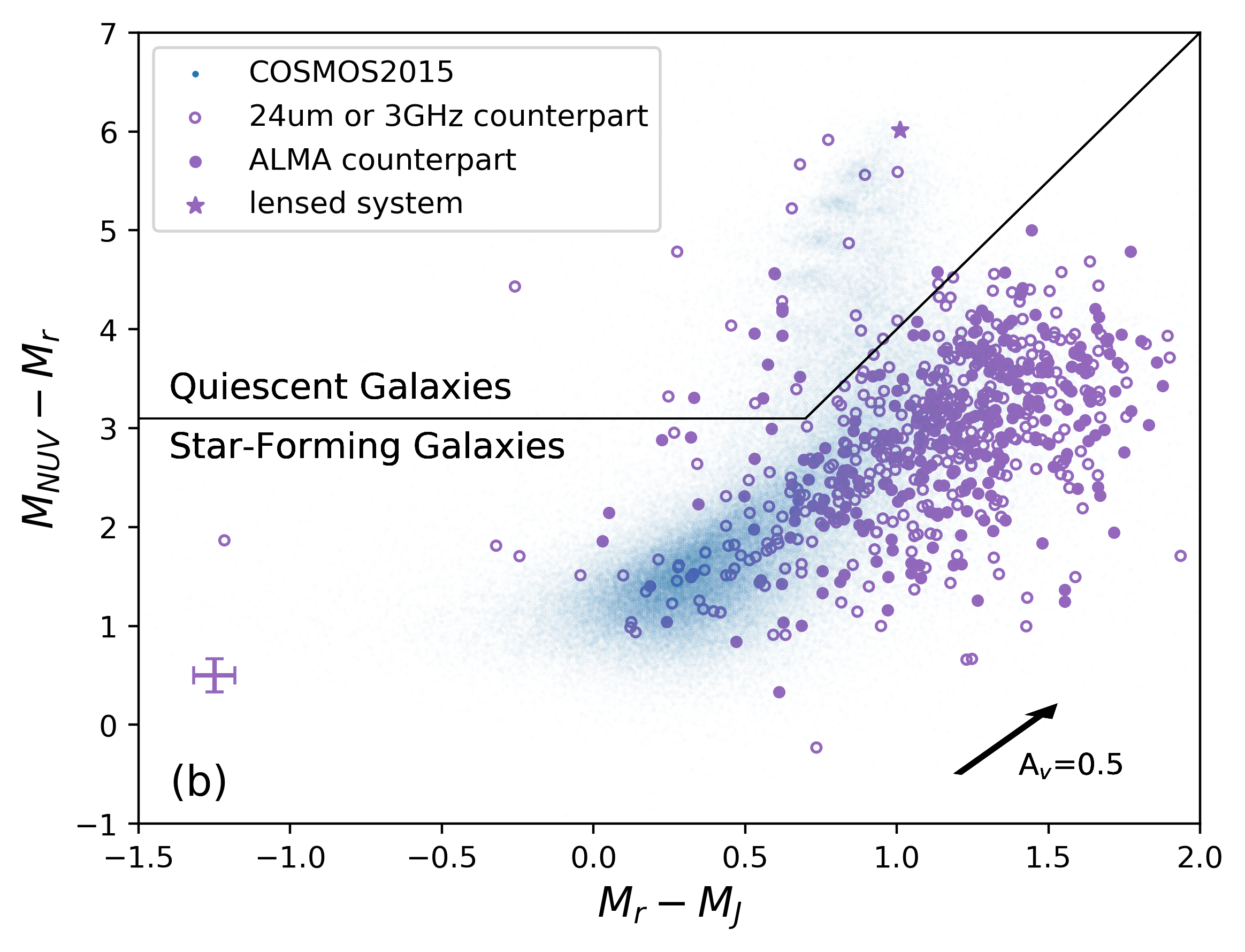}
\caption{Distributions of the 450 $\mu$m (a) and 850 $\mu$m (b) detected sample in the $NUV$--$r$--$J$ diagram. The filled circles are samples matched to ALMA sources, while the empty circles are samples matched to 24 $\mu$m or 3 GHz sources. The star symbols show the position of the lensed system described in Appendix~\ref{appendix:lensed}. The reddening vector derived from \citet{Calzetti2000} extinction and the typical errors in the two colors for the submillimeter sources are also shown with arrows and crosses, respectively.  The color errors of the 850 $\mu$m sources are larger because these sources are generally fainter in the optical than 450 $\mu$m sources. \label{fig:NUVrJ_submm}}
\end{figure}

The results of the cross-matching are summarized in Table \ref{tab:data}. For the SCUBA-2 450 $\mu$m sources, we matched 58 COSMOS2015 galaxies through ALMA observations and 181 through the MIPS and VLA catalogs. We defined them as 450 $\mu$m detected galaxies. Two out of the 58 galaxies and six out of the 181 galaxies are selected as QG candidates in the $NUV$--$r$--$J$ diagram (Fig.~\ref{fig:NUVrJ_submm} (a)). For the SCUBA-2 850 $\mu$m sources, there are 289 and 364 matches when using the ALMA catalogs and using the MIPS and VLA catalogs. We defined them as 850 $\mu$m detected galaxies. 11 out of the 289 galaxies and 19 out of the 364 galaxies are selected as QG candidates in the $NUV$--$r$--$J$ diagram (Fig.~\ref{fig:NUVrJ_submm} (b)).

One thing worth noting is the distribution of 450 $\mu$m and 850 $\mu$m detected QG candidates in the $NUV$--$r$--$J$ diagrams in Fig.~\ref{fig:NUVrJ_submm}. Although the sample sizes are small here, these QG candidates do not appear to have a tendency of locating near the selection boundaries (cf.\ the 24 $\mu$m case in Fig.~\ref{fig:NUVrJ243} (a)), comparing to the typical color errors (crosses in Fig.~\ref{fig:NUVrJ_submm}). This suggests that most of them are systems that consist of a quiescent component that dominates the rest-frame UV/optical emission and a dusty component that shows up in the far-IR. This can be either an interacting system like the one in \citet{Simpson2017} and \citet{Schreiber2018a}, or a foreground quiescent galaxy which lenses a background dusty galaxy. Indeed, one of the QG candidate is matched to both 450 and 850 $\mu$m sources through ALMA observations. This target is likely to be a lensed system (star symbol in Fig.~\ref{fig:NUVrJ_submm}), and we will further discuss it in the end of this section and in Appendix~\ref{appendix:lensed}.

From the numbers of QG candidates that have 450 or 850 $\mu$m detections, we could estimate the fraction of the bright submillimeter galaxies among our QG candidates. The results show that 0.43$\pm$0.15\% (8/1,846) and 0.16$\pm$0.03\% (30/18,304) of our QG candidates are bright 450 $\mu$m and 850 $\mu$m sources, respectively (Table \ref{tab:data}). The fraction of 450 $\mu$m detected QGs is slightly ($\sim1.8\sigma$) larger than that of 850 $\mu$m ones. This may be a result of either the better luminosity sensitivity or the higher source density at 450 $\mu$m.  The former allows us to detect more QGs at 450 $\mu$m, while the latter increases the probability of chance projection between unrelated QGs and 450 $\mu$m sources. If we remove the expected number of chance projections (Section~\ref{subsubsec:chance_projection}), then the difference reduces to $\sim1.3\sigma$. So the reason of the difference between the fractions at 450 and 850 $\mu$m remains unclear under our sample sizes.

\begin{figure}[ht!]
\epsscale{1.15}
\plotone{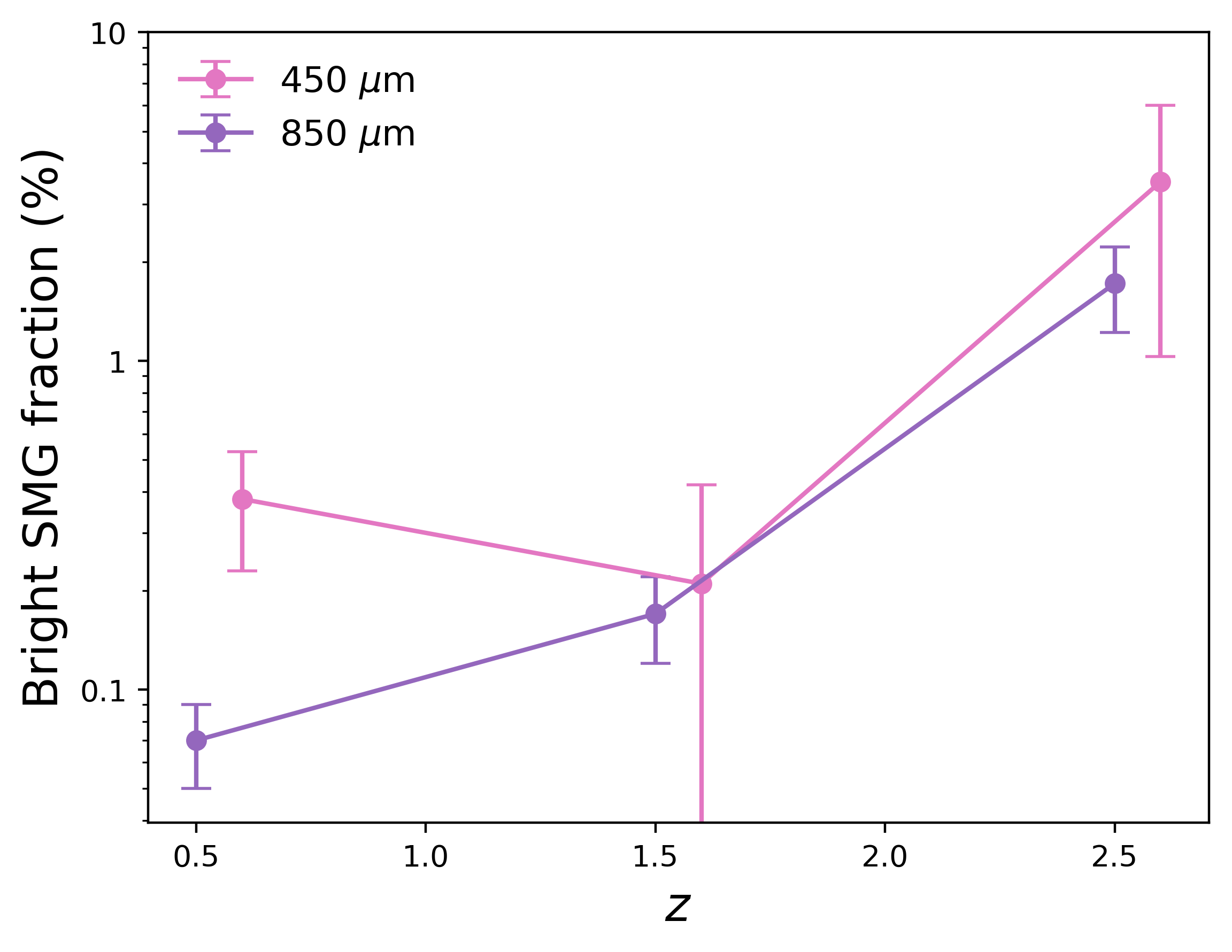}
\caption{Percentage of bright submillimeter galaxies (450 and 850 $\mu$m detected QGs) among COSMOS2015 QGs (Table \ref{tab:brightSMG}) in  logarithmic scale. The data points of 450 $\mu$m detected QGs are slightly offset along $x$-axis for clarity. The error bars of the 450 $\mu$m detected QGs are larger because of the smaller coverage of the STUDIES map. The errors are Poissonian.\label{fig:BSMGfraction}}
\end{figure}

We further spilt the populations into redshift bins (Table~\ref{tab:brightSMG} and Fig.~\ref{fig:BSMGfraction}). We can see that the fraction of 850 $\mu$m detected QG candidates increases with redshift and rises up to 3.51$\pm$2.48\% at $z>$ 2. This higher contamination rate at $z>2$ could come from either a real redshift evolution, or simply larger photometric uncertainties on high-redshift sources. Nevertheless, this few-percent contamination rate is still quite low. In conclusion, our QG candidates could be contaminated by bright dusty SFGs at a 0.16\% to 0.43\% level, and the contamination rises up to $\sim$ 1.7\% to 3.5\% at higher redshift. We note that the contamination rates may be underestimated since we may not pick up all SCUBA-2 source counterparts in the two-step counterpart identification. We will perform a ``blind'' cross-matching in Section \ref{subsec:blind_matching} to provide a different estimate of the contamination.

We analyze the role of photometric errors in the bright SMG contamination. In Section~\ref{sec:QG_selection}, we estimated the probability of intrinsically being in the SFG color space but scattered into the QG color space by photometric errors for each QG candidate. The mean probabilities are 9.6\% and 6.2\% for 450 and 850 $\mu$m detected QGs, respectively. These both account for less than 10\% of the SMG contaminations. Therefore, the bright SMG contamination is mainly due to intrinsic properties of the QGs rather than photometric errors.

\begin{deluxetable*}{l|ccc|ccc}
\tablecaption{\label{tab:brightSMG}Percentage of bright submillimeter galaxies (sub-mm detected QGs) among COSMOS2015 QGs.}
\tablehead{
\colhead{} & \colhead{total in the} & \colhead{850 $\mu$m detected} & \colhead{percentage} & \colhead{total in the} & \colhead{450 $\mu$m detected} & \colhead{percentage} \\
\colhead{} & \colhead{COSMOS field} & \colhead{} & \colhead{} & \colhead{STUDIES field} & \colhead{} & \colhead{}
}
\startdata
all          & 18304 & 30 & 0.16$\pm$0.03 \% & 1846 & 8 & 0.43$\pm$0.15 \% \\
$z\leq$ 1    & 11562 & 8  & 0.07$\pm$0.02 \% & 1314 & 5 & 0.38$\pm$0.17 \% \\
1$< z\leq$ 2 & 6045  & 10 & 0.17$\pm$0.05 \% & 475  & 1 & 0.21$\pm$0.21 \% \\
$z>$ 2       & 697   & 12 & 1.72$\pm$0.50 \% & 57   & 2 & 3.51$\pm$2.48 \% \\
\enddata
\tablecomments{The errors are set to be Poissonian. The 850 $\mu$m and 450 $\mu$m detected samples are determined through both the low-resolution SCUBA-2 data and the high resolution auxiliary data.}
\end{deluxetable*}

In the above, we used both the AS2COSMOS and the A3COSMOS catalogs during the cross-matching. We can also estimate the contamination by matching QG candidates to only the AS2COSMOS catalog, which contains a homogeneous selection and complete observations of SCUBA-2 850 $\mu$m sources with $S_{850 \mu m}>$ 6.2 mJy in the S2COSMOS map. If we match our QG candidates to 850 $\mu$m sources through only the AS2COSMOS catalog, we found 7 galaxies to be 850 $\mu$m detected QG candidates. If we assume the same QG fraction for all SCUBA-2 850 $\mu$m sources, we estimate that there should be 36.9$\pm$14.0 QG candidates. This accounts for 0.2$\pm$0.1\% among all the QG candidates. This agrees with the 0.16$\pm$0.03\% contamination mentioned above.  

Furthermore, the SCUBA-2 catalog has a detection limit of 2 mJy but is not complete for sources above 2 mJy.  The complete number of 850 $\mu$m sources can be estimated from the sources counts corrected for completeness, from \citet{Simpson2019}.  If we estimate the complete number of sources above 2 mJy and assume the same QG fraction in the AS2COSMOS catalog, we obtain a dusty galaxy contamination rate of 0.6$\pm$0.3\%. The relative uncertainty here is slightly larger than that simply propagated from the number of QGs in the AS2COSMOS catalog since the source counts also contain an uncertainty. Nevertheless, this value is larger than the above-estimated value of 0.16$\pm$0.03\% and is probably a more realistic estimate if we do have a deeper and more complete survey at 850 $\mu$m.

We note that one out of the 11 QG candidates (ID = 659416 in the COSMOS2015 catalog) that are matched to SCUBA-2 850 $\mu$m sources through ALMA catalogs is likely to be a lensed system because of its unusual submillimeter/radio flux ratio (see Appendix~\ref{appendix:lensed} for details). This example demonstrates that when matching QG candidates to the submillimeter sources, a match does not imply the QG candidate and the long-wavelength source to be the same object. They could be physical associations such as the lensed system here, or an interacting galaxy pair consisting of a QG and a dusty object \citep[e.g.,][]{Schreiber2018a}. Based on our small sample size, the probability for such association is about 9\% (1/11). Such spatial correlation effects caused by lensing or galaxy interaction will be further discussed in Section \ref{subsec:blind_matching}.

\subsubsection{Effect of Chance Projection}
\label{subsubsec:chance_projection}

Given the small numbers of matched objects in the previous section, we would like to examine whether the matches between our QG candidates and bright submillimeter sources are caused by chance projection or by real spatial correlation. We could estimate the effect of chance projection by simple calculations.

First, we calculated the search area in our 2-step cross matching. For the SCUBA-2 sources with ALMA observation, we used a search radius of $1\arcsec$ for the ALMA sources. For the remaining ones, we used search radii of $2\arcsec$ and $0.8\arcsec$ for 24 $\mu$m and 3 GHz sources, respectively. If a 3 GHz source located within the $2\arcsec$ search radius of a 24 $\mu$m source, we only adapted the search area of the 3 GHz source. We then calculated the expected fraction of randomly distributed QG candidates locating in the search area with $1-e^{-na/A}$, where $A$ is the survey area, $n$ is the number of searched sources in the high-resolution catalogs, and $a$ is the search area per source. The estimated numbers of chance projections are given in the last column of Table \ref{tab:data}. When we matched through ALMA catalogs, the numbers of chance projections is significantly lower. When we matched through 24 $\mu$m and 3GHz catalogs, the probability of chance projections is about 1/3; 2.1 out of 6 (35.0\%) and 5.8 out of 19 (30.5\%) matches can be chance projections for 450 and 850 $\mu$m detected QG candidates.

We conclude that among the submillimeter detected QG candidates mentioned in the previous section, accounting for 0.16\% to 0.43\% among our QG candidates, the majority are real physical associations. The estimated bright dusty SFG contamination is not mainly driven by chance projections.

\subsection{Blind Cross-Matching} \label{subsec:blind_matching}

\subsubsection{Matching with Large Radii and Estimate of Chance Projection}

In the cross-matching aided by 3 GHz and 24 $\mu$m astrometry described in Secton \ref{subsec:traditional_matching}, there is a possibility that the real optical counterparts of the submillimeter sources are undetected at 3 GHz and/or 24 $\mu$m. The different redshift dependences of sensitivities in the submillimeter, radio, and mid-IR may introduce such a bias.  To avoid this, we can perform a ``blind'' cross-matching to the SCUBA-2 sources. We directly match the QG candidates with SCUBA-2 450 $\mu$m and 850 $\mu$m sources using $4\arcsec$ and $7\arcsec$ search radii, respectively, without relying on radio and mid-IR positions. The large matching radii here will unavoidably lead to larger numbers of chance projections, so we need to more precisely estimate the number of chance projections.

\begin{figure}[ht!]
\epsscale{1.15}
\plotone{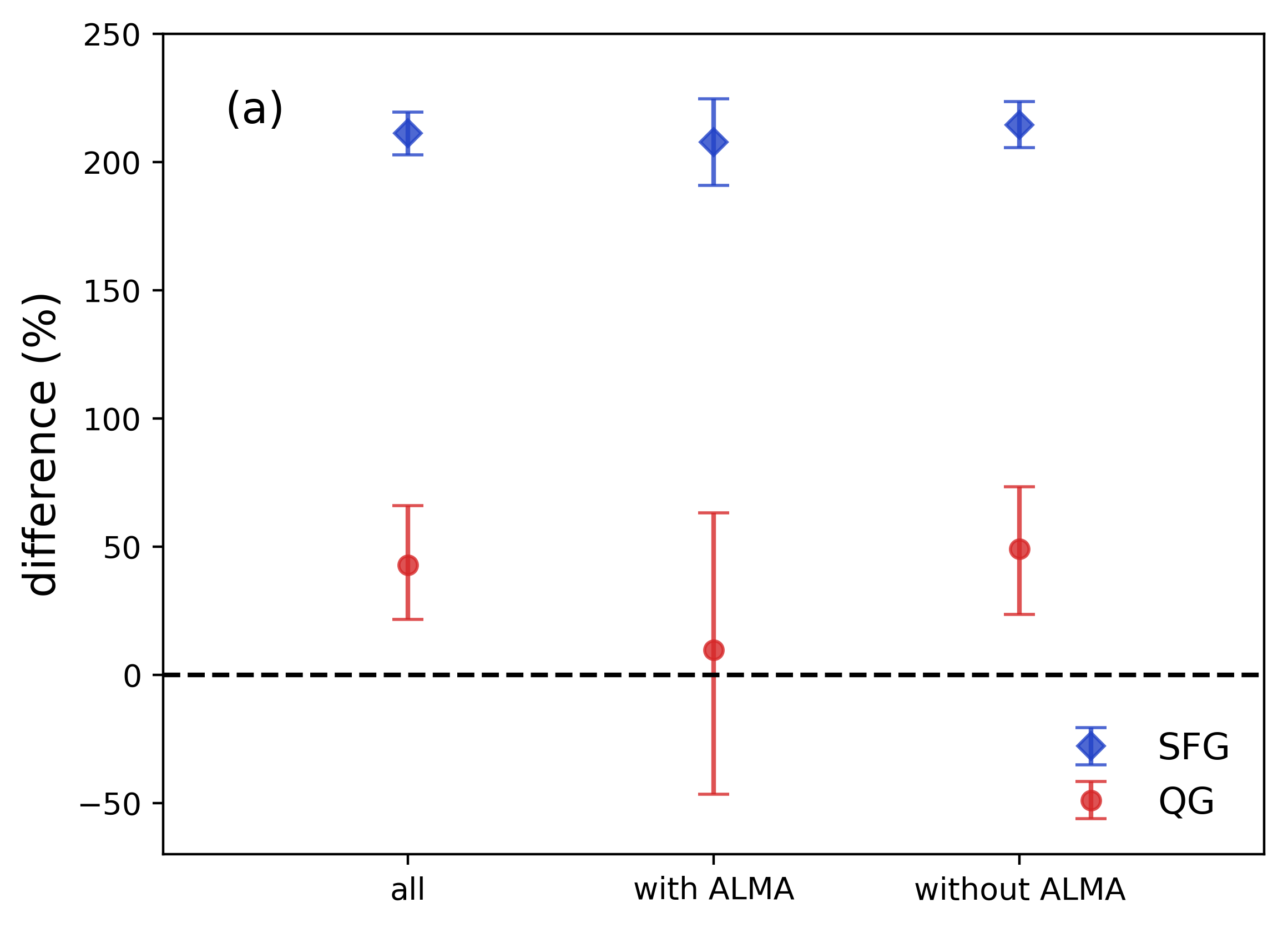}
\plotone{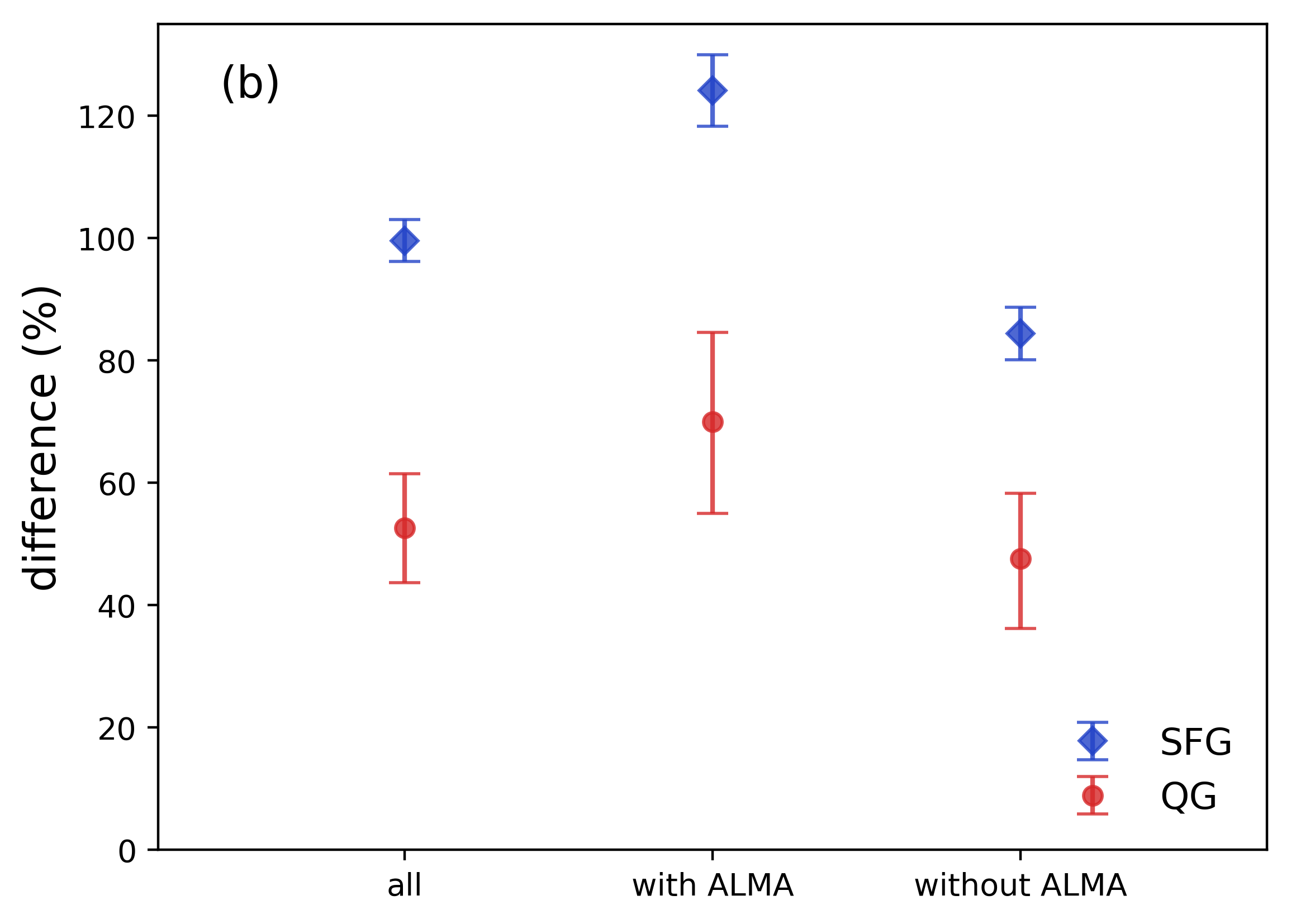}
\plotone{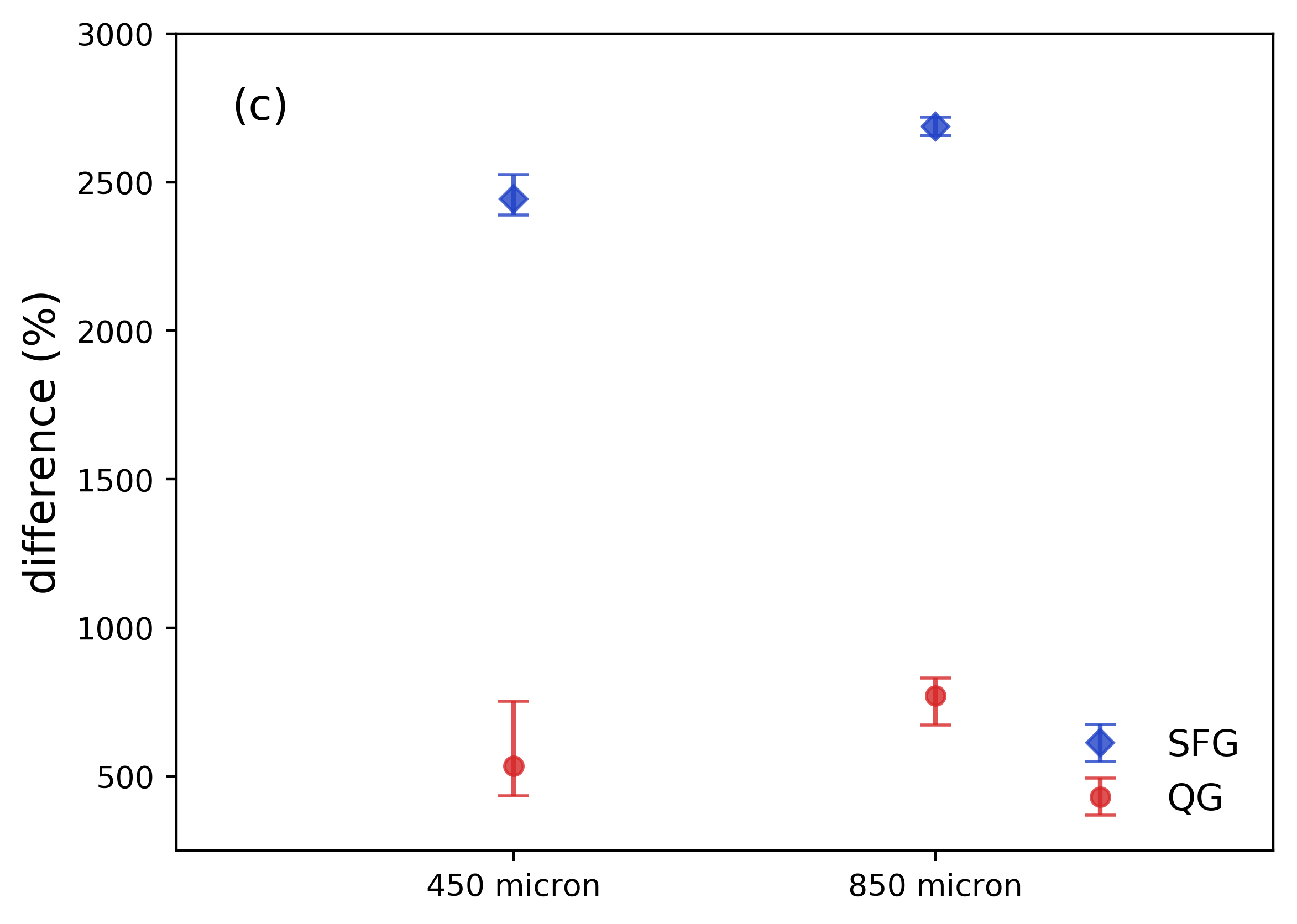}
\caption{Fractional differences between the actual matches and the expected matches based on random spatial distributions (see also Table~\ref{tab:spatial}). Panels (a), (b), and (c) show results for SCUBA-2 450 $\mu$m, 850 $\mu$m, and ALMA sources, where $4\arcsec'$, $7\arcsec$, and $1\arcsec$ search radii are used respectively. It can be seen that the QG results for the SCUBA-2 850 $\mu$m sources and ALMA sources are all significantly above zero, showing that these extra matches in the actual sample are caused by real physical connection between the QG candidates and the submillimeter sources, rather than chance projection.\label{fig:spatial_plot}}
\end{figure}

To do this, we simulated the matching results using SCUBA-2 submillimeter sources with random positions. Here we do not apply the $1-e^{-na/A}$ method because the distribution of QGs may not be random at the scale of the relatively large search radii for SCUBA-2 sources and therefore the dispersion in the mean cannot be estimated.  We calculated the expected number of QG candidates located within a search radius from the randomly distributed submillimeter sources and compared the results with the actual number of matched QG candidates. The simulation is repeated 1,000 times. The estimated number of matches and its error are set to be the mean and the 68\% interval of the 1,000 results.  The results are summarized in Table \ref{tab:spatial}, and the fractional difference between the expected matches (chance projections) and the actual matches are also shown in Fig. \ref{fig:spatial_plot}. We note that the detection limit of the SCUBA-2 450 $\mu$m, 850 $\mu$m, and ALMA sources are different. We also estimated the probability that the expected number is equal to or larger than the actual number (Table \ref{tab:spatial}). We show the results of SFGs for comparison in Table \ref{tab:spatialSFG}.

\begin{deluxetable*}{cccc|ccccc}
\tablecaption{Cross-Matches and Expected Chance Projections between QGs and Submillimeter Sources \label{tab:spatial}}
\tablehead{
\colhead{SCUBA-2} & \colhead{Group} & \colhead{Match} & \colhead{Number}
& \colhead{Expected} & \colhead{Actual}
& \colhead{Difference} & \colhead{Fractional} & \colhead{Probability\tablenotemark{b}}\\
\colhead{sources} & \colhead{} & \colhead{Radius\tablenotemark{a}} & \colhead{}
& \colhead{Matched QG} & \colhead{Matched QG}
& \colhead{} & \colhead{Difference} & \colhead{}
}
\startdata
           & all          & $4\arcsec$  & 353 & 
20.3$^{+ 4.7 }_{- 4.3 }$ &  29  &  8.7$^{+ 4.7 }_{- 4.3 }$ &  42.6$^{+ 22.9 }_{- 21.3 }$\% & 0.05 \\
450 $\mu$m & without ALMA & $4\arcsec$  & 276 & 
16.1$^{+ 3.9 }_{- 4.1 }$ &  24  &  7.9$^{+ 3.9 }_{- 4.1 }$ &  48.9$^{+ 24.1 }_{- 25.5 }$\% & 0.046 \\
sources    & with ALMA    & $4\arcsec$  & 77  &
4.6$^{+ 2.4 }_{- 2.6 }$ &  5  &  0.4$^{+ 2.4 }_{- 2.6 }$ &  9.6$^{+ 53.5 }_{- 56.1 }$\% & 0.475 \\
           & ALMA sources & $1\arcsec$  & 85  &
0.3$^{+ 0.7 }_{- 0.3 }$ &  2  &  1.7$^{+ 0.7 }_{- 0.3 }$ &  534.9$^{+ 217.5 }_{- 100.0 }$\% & 0.042 \\
\hline
           & all          & $7\arcsec$  & 981 &
135.0$^{+ 12.0 }_{- 12.0 }$ &  206  &  71.0$^{+ 12.0 }_{- 12.0 }$ &  52.6$^{+ 8.9 }_{- 8.9 }$\% & 0 \\
850 $\mu$m & without ALMA & $7\arcsec$  & 611 &
84.6$^{+ 10.4 }_{- 9.6 }$ &  120  &  35.4$^{+ 10.4 }_{- 9.6 }$ &  41.8$^{+ 12.2 }_{- 11.4 }$\% & 0 \\
sources    & with ALMA    & $7\arcsec$  & 370 &
50.6$^{+ 7.4 }_{- 7.6 }$ &  86  &  35.4$^{+ 7.4 }_{- 7.6 }$ &  69.8$^{+ 14.5 }_{- 15.1 }$\% & 0 \\
           & ALMA sources & $1\arcsec$  & 452 &
1.3$^{+ 0.7 }_{- 1.3 }$ &  11  &  9.7$^{+ 0.7 }_{- 1.3 }$ &  771.6$^{+ 58.5 }_{- 100.0 }$\% & 0 \\
\enddata
\tablenotetext{a}{The radius of $1\arcsec$ to $7\arcsec$ correspond to 8--56 kpc at $z=1$ and 8--57 kpc at $z=2.5$.}
\tablenotetext{b}{The probability that the expected number of matches (based on random spatial distribution) is equal to or larger than that of the actual matches.}
\end{deluxetable*}

\begin{deluxetable*}{cccc|ccccc}
\tablecaption{Cross-Matches and Expected Chance Projections between SFGs and Submillimeter Sources \label{tab:spatialSFG}}
\tablehead{
\colhead{SCUBA-2} & \colhead{Group} & \colhead{Match} & \colhead{Number}
& \colhead{Expected} & \colhead{Actual}
& \colhead{Difference} & \colhead{Fractional} & \colhead{Probability\tablenotemark{b}}\\
\colhead{sources} & \colhead{} & \colhead{Radius\tablenotemark{a}} & \colhead{}
& \colhead{Matched SFG} & \colhead{Matched SFG}
& \colhead{} & \colhead{Difference} & \colhead{}
}
\startdata
           & all          & $4\arcsec$  & 353 & 
149.7$^{+ 12.3 }_{- 12.7 }$ &  466  &  316.3$^{+ 12.3 }_{- 12.7 }$ &  211.3$^{+ 8.2 }_{- 8.5 }$\% & 0 \\
450 $\mu$m & without ALMA & $4\arcsec$  & 276 & 
116.4$^{+ 10.6 }_{- 10.4 }$ &  366  &  249.6$^{+ 10.6 }_{- 10.4 }$ &  214.3$^{+ 9.1 }_{- 9.0 }$\% & 0 \\
sources    & with ALMA    & $4\arcsec$  & 77  &
32.5$^{+ 5.5 }_{- 5.5 }$ &  100  &  67.5$^{+ 5.5 }_{- 5.5 }$ &  207.6$^{+ 16.9 }_{- 16.9 }$\% & 0 \\
           & ALMA sources & $1\arcsec$  & 85  &
2.2$^{+ 1.8 }_{- 1.2 }$ &  56  &  53.8$^{+ 1.8 }_{- 1.2 }$ &  2443.1$^{+ 81.7 }_{- 54.6 }$\% & 0 \\
\hline
           & all          & $7\arcsec$  & 981 &
1045.7$^{+ 35.5 }_{- 35.9 }$ &  2087  &  1041.3$^{+ 35.5 }_{- 35.9 }$ &  99.6$^{+ 3.4 }_{- 3.4 }$\% & 0 \\
850 $\mu$m & without ALMA & $7\arcsec$  & 611 & 
652.9$^{+ 28.1 }_{- 27.9 }$ &  1204  &  551.1$^{+ 28.1 }_{- 27.9 }$ &  84.4$^{+ 4.3 }_{- 4.3 }$\% & 0 \\
sources    & with ALMA    & $7\arcsec$  & 370 &
394.0$^{+ 23.0 }_{- 23.0 }$ &  883  &  489.0$^{+ 23.0 }_{- 23.0 }$ &  124.1$^{+ 5.8 }_{- 5.8 }$\% & 0 \\
           & ALMA sources & $1\arcsec$  & 452 &
9.9 $^{+ 3.1 }_{- 2.9 }$ &  277  &  267.1$^{+ 3.1 }_{- 2.9 }$ &  2687.3$^{+ 30.8 }_{- 29.6 }$\% & 0 \\
\enddata
\tablenotetext{ab}{~The parameters follow those in Table \ref{tab:spatial}.}
\end{deluxetable*}

The first and fifth rows of Table \ref{tab:spatial} and \ref{tab:spatialSFG} presents simple cross-matching between the QGs/SFGs and single-dish submillimeter samples without information from ALMA.  For SCUBA-2 450 $\mu$m sources, the actual matches are 42.6$^{+22.9}_{-21.3}$\% larger than the expected random matches. Although the significance is only about 2$\sigma$ (Fig. \ref{fig:spatial_plot} (a)), the estimated probability that the expected number equals to or is larger than the actual number is 0.05, which is quite low. This result suggests that there are $8.7^{+4.7}_{-4.3}$ QGs physically associated with the 353 450 $\mu$m sources. This corresponds to $0.47^{+0.25}_{-0.23}\%$ of the 1846 QGs in the 450 $\mu$m map area. The statistically derived number of $8.7^{+4.7}_{-4.3}$ matches also nicely agrees with the 8 matches found with high-resolution data (Section~\ref{subsec:traditional_matching} and Table~\ref{tab:data}). Also, for comparison, $316.3^{+12.3}_{-12.7}$ SFGs are physically associated with the 353 450 $\mu$m sources.  This is as expected, since dusty submillimeter sources should be dominated by SFGs.

For SCUBA-2 850 $\mu$m sources, the actual matches are 52.6$^{+8.9}_{-8.9} \%$ larger than the expected random matches (see also \ref{fig:spatial_plot} (b)), which is significant ($\sim6\sigma$). The estimated probability that the expected number equals to or is larger than the actual number is nearly zero. This means that among the 206 matches between the QGs and 850 $\mu$m sources, $71\pm12$ are real physical associations. These $\sim71$ sources account for $0.39\pm0.07\%$ of the 18304 QGs. We note that the number of 71 is significantly different from the number of 30 quoted in Table~\ref{tab:data}, or 23 after removing the expected number of chance projections.  This is because here we do not require high-resolution multi-wavelength data to pin down the cross-matching.  This suggests that a significant fraction of QG--850 $\mu$m associations in our data do not have 24 $\mu$m, 3 GHz, or ALMA counterparts. This is perhaps partially because of the insufficient sensitivity (24 $\mu$m and 3 GHz cases) and incomplete coverage (ALMA cases). However, we will soon show that a large fraction of these 71 sources are clustered around the SCUBA-2 sources at a $\sim7\arcsec$ scale, but they are not the submillimeter, mid-IR, or radio emitters. Finally, as in the case for 450 $\mu$m cross-matching, the overlap between 850 $\mu$m sources and SFGs is much larger than that for QG, which is expected.

\subsubsection{Verification and Comparison with ALMA Sources}

The above ``blind'' matching between QG candidates and SCUBA-2 sources using large matching radii (1/2 of the single-dish beam FWHM) and the comparison between actual matches and simulations allows us to statistically assess the numbers of real physical associations.  Now with the ALMA data, we can pin down the cross-matching with a much smaller matching radius, with a smaller subsample.  We split the SCUBA-2 sources into those with and without ALMA observations. The results are listed in the remaining rows of Table \ref{tab:spatial} and \ref{tab:spatialSFG}.

As a simple sanity check, we ran cross-matching with large matching radii over the subsamples. The fractional differences between actual and expected random matches do not change significantly between the ALMA and non-ALMA subsamples for QGs (Fig. \ref{fig:spatial_plot} (a) and (b)). This is even true for the 450 $\mu$m sample, albeit the small sample sizes and therefore the large errors.  This implies that there is no special selection bias in the ALMA observations regarding their QG-submillimeter properties.

For the SCUBA-2 sources with ALMA observations, the expected numbers for matches under a $1\arcsec$ search radius and random spatial distributions are always small comparing to the actual matches. This is reflected on the large fractional differences between the actual matches and expected random matches, which are 534.9$^{+217.5}_{-100.0}$\% for the 450 $\mu$m sources and 771.6$^{+58.5}_{-100.0}$\% for the 850 $\mu$m sources (Fig.~\ref{fig:spatial_plot}(c), the fourth and eighth rows).  This means that the majority of the observed matches between QGs and ALMA sources under a $1\arcsec$ matching radius are real physical associations.

An interesting comparison is to see if the $1\arcsec$ matching pinned down by ALMA agrees with the statistical estimates of real physical associations derived from the large-radius blind matching. Table~\ref{tab:spatial} and \ref{tab:spatialSFG} show that if we match the 77 SCUBA-2 450 $\mu$m sources to QGs and SFGs using a $4\arcsec$ matching radius, we expect $0.4^{+2.4}_{-2.6}$ out of the 5 QG matches and $67.5^{+5.5}_{-5.5}$ out of the 100 SFG matches to be real associations.  These can be compared with the ALMA results for the same sub-sample: 1.7$^{+ 0.7 }_{- 0.3 }$ and 53.8$^{+ 1.8 }_{- 1.2 }$ real associations for the QGs and the SFGs.  The values for QG--SMG associations agree nicely, albeit the small sample size.  This probably validates the statistical method for estimating the number of real associations and chance projections using simulations and random distributions.  On the other hand, the values for SFG--SMG associations (67.5 and 53.8) differ by 25\% and the difference is about $2\sigma$. The excess in the number of SFGs around SMGs within $4\arcsec$ comparing to the number of true associations pinpointed by ALMA suggests a weak clustering of SFGs around SMGs.  This excess is only $2\sigma$ and is not statistically significant. However, if we look at the 850 $\mu$m values, the excesses for both QG--SMG and SFG--SMG associations become highly significant.

We make a similar comparison on the 850 $\mu$m ALMA subsample.  The expected numbers of real associations under a $7\arcsec$ matching radius for the 370 850 $\mu$m sources are $35.4^{+7.4}_{-7.6}$ and $489.0^{+23.0}_{-23.0}$ for QGs and SFGs, respectively. However, the numbers revealed by ALMA observations are much smaller: 9.7$^{+ 0.7 }_{- 1.3 }$ and 267.1$^{+ 3.1 }_{- 2.9 }$.  The differences between the two sets of numbers are both significant.  This implies that once we increase the matching radius from $1\arcsec$ to $7\arcsec$ ($\lesssim 60$ kpc at $z=1$--2), additional clustering effects kick in, i.e., there are QGs and SFGs physically associated with the submillimeter sources at such large scales, but they are not the submillimeter sources themselves nor arcsec-scale galaxy-galaxy lensing pairs. This effect becomes undetectable (QGs) or much weaker (SFGs) under the $4\arcsec$ matching for the 450 $\mu$m sources, either because of the small sample sizes for the 450 $\mu$m analysis or because of the different spatial distribution for low-dust-luminosity sources.

Previous studies of QG autocorrelation functions found that QGs show stronger clusting signal than SFGs at arcminute scales \citep{Williams2009}, but there do not exist QG-SFG or QG-SMG cross-correlation analyses. Our results suggest that QGs and SMGs are clustered, and detailed cross-correlation studies between these two distinct populations will be an interesting future topic.

In summary, with direct cross-matching to SCUBA-2 sources and statistical analyses of chance projection effects, we do not find evidence for a different dusty galaxy contamination rate among QGs comparing to what we found with counterpart identifications using ALMA, 24 $\mu$m, and 3 GHz data. Instead, we found a clustering effect between the bright submillimeter sources and our QG candidates at scales from $1\arcsec$ to $7\arcsec$ ($\sim8$--60 kpc at $z=1$--2).

We note that our studies in Section \ref{subsubsec:traditional_matching_process} and \ref{subsec:blind_matching} thus far imply several possibilities for the submillimeter detected QG candidates obtained from the cross-matching process in Section \ref{subsubsec:traditional_matching_process}, depending on the angular scales to which the observations are sensitive. They could be the correct submillimeter counterparts to the QG candidates. There are also situations where the QG candidates are not submillimeter emitters, but are physically associated with submillimeter galaxies through effects like galaxy-galaxy lensing (e.g., Fig.~\ref{fig:lensing}), galaxy interaction, or clustering effects at scale of a few arcsec.

\section{Faint Submillimeter Galaxies Among QG Candidates} \label{sec:faint_SMG}

In Section \ref{sec:bright_SMG}, we matched our QGs candidates to submillimeter sources and demonstrated that fractions of the matched QG candidates are physically related to the submillimeter sources. However, the 450 $\mu$m and 850 $\mu$m sources have a detection limit of about 3.5 mJy and 2 mJy, respectively, which correspond to SFR of roughly 60 and 180 $M_{\odot}$ year$^{-1}$ at $z = 1$. Therefore, we further perform stacking analysis in order to search for fainter submillimeter emissions among the QG candidates.

\subsection{Stacking Analysis}

We measured the submillimeter emission from the SCUBA-2 maps at the positions of our selected QG candidates and calculated the error-weighted average of their fluxes. As our sources are point-like under JCMT's resolution and the SCUBA-2 maps were beam-matched to produce maximum-likelihood flux for point sources, fluxes are measured by directly reading the map values in Jy beam$^{-1}$ at the positions of the QGs. We excluded QG candidates that we matched to the bright submillimeter sources in Section \ref{subsec:traditional_matching}, as well as QG candidates whose measured SCUBA-2 fluxes exceed $3\sigma$, in order to prevent our results from being biased by the small number of bright submillimeter sources. To estimate the bias and uncertainty in such a stacked flux, we then stacked at 1,000 random positions and repeated this 10,000 times. In this process, bright submillimeter sources are removed according to the same criteria as above.  The mean from these random stacks is considered as the bias in stacking. It is consistent with zero, because of the zero-sum nature of the match-filtered SCUBA-2 maps. Nevertheless, this small bias is subtracted from the mean of the QGs. The dispersion among the 10,000 measurements of the random samples is considered as the uncertainty of stacking 1,000 sources. It is scaled by 1/$\sqrt{N}$ to be the uncertainty of the QG stacking.  We also stacked different numbers of random sources to verify this 1/$\sqrt{N}$ dependence.

The stacking results are shown in Table \ref{tab:450stack} and Table \ref{tab:850stack}.  The first rows of the two tables show that we can reach a $6.3\sigma$ statistical detection at 850 $\mu$m if we simply stack all QG candidates, but not a significant detection at 450 $\mu$m. The non-detection at 450 $\mu$m may be due to the smaller coverage of the STUDIES map.  Furthermore, we can divide the QG candidates into subgroups according to their properties, to see if there is a particular group of QGs that contributes to the majority of the stacked signal. First, we classified QG candidates either with 24 $\mu$m counterparts or with 3 GHz counterparts labeled with SFG flags in the VLA catalog \citep{Smolcic2017} as ``IR-radio-bright'' QGs, and the rests as ``IR-radio-faint'' QGs. Our terminology is similar but slightly different from that in \citet{Man2016}. \citet{Man2016} defined QG candidates with SFR derived from 24 $\mu$m over 100 $M_{\odot}$ year$^{-1}$ as ``IR-bright'' QGs, and the rests as ``IR-faint'' QGs. They used 24 $\mu$m data and SFR constraints to classify the subgroups, while we used 24 $\mu$m data, 3 GHz data, and radio AGN classification in our work. Then, for the 850 $\mu$m stacking, because of the larger area of the SCUBA-2 map and therefore more available QGs, we can further divide the QG sample into various redshift and stellar mass bins.

\begin{deluxetable*}{lcccccccc}
\tablecaption{450 $\mu$m QG Stacking Results\label{tab:450stack}}
\tablehead{
\colhead{Groups} & \colhead{log$M_*$\tablenotemark{a}} & \colhead{$z$\tablenotemark{b}} & 
\colhead{Number} & \colhead{$S_{450\rm \mu m}$} & \colhead{SNR}      &
\colhead{log($L_{\rm IR}$)} & \colhead{SFR$_{450\rm \mu m}$} & \colhead{SFR$_{\rm optical}$\tablenotemark{c}}\\
& \colhead{(log($M_{\odot}$))} & \colhead{} & \colhead{} & \colhead{(mJy)} & \colhead{} &  \colhead{(log($L_{\odot}$))} &  \colhead{($M_{\odot}$ yr$^{-1}$)} &  \colhead{($M_{\odot}$ yr$^{-1}$)}
}
\startdata
All QG                 & 10.7$^{+0.3}_{-1.1}$ & 0.9 & 1799 & 0.06$\pm$0.05 & 1.3 & 9.3$^{+0.3}_{-0.7}$  & 0.2$\pm$0.2 & 2.5$^{+0.0}_{-0.0}$  \\
24-$\mu$m counterpart  & 10.9$^{+0.3}_{-0.7}$ & 0.8 & 155  & 0.66$\pm$0.16 & 4.1 & 10.6$^{+0.1}_{-0.1}$ & 3.6$\pm$0.9 & 11.2$^{+0.4}_{-0.3}$ \\
3-GHz counterpart      & 11.2$^{+0.2}_{-0.4}$ & 0.9 & 103  & 0.60$\pm$0.20 & 3.0 & 10.5$^{+0.1}_{-0.2}$ & 3.5$\pm$1.2 & 9.5$^{+0.4}_{-0.4}$  \\
3-GHz counterpart: SFG & 11.1$^{+0.2}_{-0.4}$ & 0.9 & 45   & 0.85$\pm$0.30 & 2.8 & 10.8$^{+0.1}_{-0.2}$ & 6.4$\pm$2.3 & 10.3$^{+0.8}_{-0.7}$ \\
3-GHz counterpart: AGN & 11.2$^{+0.2}_{-0.3}$ & 0.9 & 58   & 0.39$\pm$0.26 & 1.5 & 10.2$^{+0.2}_{-0.5}$ & 1.5$\pm$1.0 & 8.9$^{+0.5}_{-0.6}$  \\
IR-radio-faint QG      & 10.6$^{+0.3}_{-1.1}$ & 0.9 & 1620 & 0.00$\pm$0.05 & 0.0 & 0.0$^{+9.3}_{-0.0}$  & 0.0$\pm$0.2 & 1.7$^{+0.0}_{-0.0}$  \\
IR-radio-bright QG     & 11.0$^{+0.3}_{-0.6}$ & 0.8 & 179  & 0.65$\pm$0.15 & 4.3 & 10.6$^{+0.1}_{-0.1}$ & 3.6$\pm$0.8 & 10.2$^{+0.4}_{-0.2}$ \\
\enddata
\tablenotetext{a}{Mean and 68\% interval of stellar mass in logarithmic scale.}
\tablenotetext{b}{Median of redshift.}
\tablenotetext{c}{Mean of SFRs from COSMOS2015. The error shows the typical error in COSMOS2015 scaled by 1/$\sqrt{N}$. Uncertainty of template fitting is not included, which may be large for QG population.}
\end{deluxetable*}

\begin{deluxetable*}{lcccccccc}
\tablecaption{850 $\mu$m QG Stacking Results\label{tab:850stack}}
\tablehead{
\colhead{Groups} & \colhead{log$M_*$\tablenotemark{a}} & \colhead{$z$\tablenotemark{b}} & 
\colhead{Number} & \colhead{$S_{850\rm \mu m}$} & \colhead{SNR}      &
\colhead{log($L_{\rm IR}$)} & \colhead{SFR$_{850\rm \mu m}$} & \colhead{SFR$_{\rm optical}$\tablenotemark{c}}\\
& \colhead{(log($M_{\odot}$))} & \colhead{} & \colhead{} & \colhead{(mJy)} & \colhead{} &  \colhead{(log($L_{\odot}$))} &  \colhead{($M_{\odot}$ yr$^{-1}$)} &  \colhead{($M_{\odot}$ yr$^{-1}$)}
}
\startdata
All QG                 & 10.7$^{+0.3}_{-1.0}$ & 0.9 & 18011 & 0.06$\pm$0.01 & 6.3 & 10.0$^{+0.1}_{-0.1}$ & 1.0$\pm$0.2  & 3.0$^{+0.0}_{-0.0}$\\
24-$\mu$m counterpart  & 10.9$^{+0.3}_{-0.6}$ & 0.8 & 1538  & 0.27$\pm$0.03 & 8.2 & 11.1$^{+0.0}_{-0.1}$ & 11.7$\pm$1.4 & 6.2$^{+0.1}_{-0.1}$\\
3-GHz counterpart      & 11.1$^{+0.2}_{-0.4}$ & 0.9 & 1028  & 0.14$\pm$0.04 & 3.5 & 10.8$^{+0.1}_{-0.1}$ & 6.8$\pm$1.9  & 6.1$^{+0.1}_{-0.1}$\\
3-GHz counterpart: SFG & 11.1$^{+0.2}_{-0.5}$ & 0.9 & 473   & 0.30$\pm$0.06 & 5.0 & 11.1$^{+0.1}_{-0.1}$ & 13.9$\pm$2.7 & 7.6$^{+0.2}_{-0.1}$\\
3-GHz counterpart: AGN & 11.2$^{+0.2}_{-0.4}$ & 0.9 & 555   & 0.01$\pm$0.05 & 0.1 & 9.2$^{+1.0}_{-9.2}$  & 0.2$\pm$1.4  & 4.9$^{+0.1}_{-0.1}$\\
\hline
IR-radio-faint QG      & 10.7$^{+0.3}_{-1.0}$ & 0.9 & 16242 & 0.04$\pm$0.01 & 3.8 & 9.8$^{+0.1}_{-0.1}$ & 0.7$\pm$0.2 & 2.6$^{+0.0}_{-0.0}$\\
~~~$z\leq$ 0.5         &&&&&&&\\
~~~log$M_*\leq$ 10.5   & 9.8$^{+0.4}_{-1.6}$  & 0.3 & 2399  &-0.01$\pm$0.03 &-0.4 & 0.0$^{+9.3}_{-0.0}$ & 0.0$\pm$0.2 & 0.0$^{+0.0}_{-0.0}$\\
~~~log$M_*>$ 10.5      & 10.9$^{+0.1}_{-0.3}$ & 0.4 & 632   & 0.07$\pm$0.05 & 1.4 & 9.8$^{+0.2}_{-0.6}$ & 0.7$\pm$0.5 & 0.1$^{+0.0}_{-0.0}$\\
~~~0.5 $< z\leq$ 1.0   &&&&&&&\\
~~~log$M_*\leq$ 10.5   & 10.1$^{+0.3}_{-0.6}$ & 0.8 & 3739  & 0.01$\pm$0.02 & 0.4 & 9.2$^{+0.6}_{-9.2}$ & 0.2$\pm$0.4 & 0.4$^{+0.0}_{-0.0}$\\
~~~log$M_*>$ 10.5      & 10.9$^{+0.2}_{-0.3}$ & 0.8 & 3375  & 0.05$\pm$0.02 & 2.2 & 9.9$^{+0.2}_{-0.3}$ & 0.8$\pm$0.4 & 0.3$^{+0.0}_{-0.0}$\\
~~~1.0 $< z\leq$ 1.5   &&&&&&&\\
~~~log$M_*\leq$ 10.5   & 10.2$^{+0.2}_{-0.3}$ & 1.2 & 1461  &-0.02$\pm$0.03 &-0.5 & 0.0$^{+9.8}_{-0.0}$ & 0.0$\pm$0.6 & 2.1$^{+0.0}_{-0.0}$\\
~~~log$M_*>$ 10.5      & 10.9$^{+0.2}_{-0.3}$ & 1.2 & 2351  & 0.09$\pm$0.03 & 3.4 & 10.6$^{+0.1}_{-0.2}$ & 3.6$\pm$1.0 & 1.1$^{+0.0}_{-0.0}$\\
~~~1.5 $< z\leq$ 2.0   &&&&&&&\\
~~~log$M_*\leq$ 10.5   & 10.3$^{+0.2}_{-0.2}$ & 1.7 & 469   & 0.07$\pm$0.06 & 1.3 & 10.3$^{+0.3}_{-0.7}$ & 2.1$\pm$1.7 & 12.6$^{+0.5}_{-0.3}$\\
~~~log$M_*>$ 10.5      & 10.9$^{+0.2}_{-0.3}$ & 1.7 & 1234  & 0.08$\pm$0.04 & 2.3 & 10.6$^{+0.2}_{-0.2}$ & 3.7$\pm$1.6 & 8.9$^{+0.1}_{-0.1}$\\
~~~2.0 $< z\leq$ 2.5   &&&&&&&\\
~~~log$M_*\leq$ 10.5   & 10.3$^{+0.2}_{-0.2}$ & 2.3 & 107   & 0.11$\pm$0.12 & 0.9 & 10.5$^{+0.3}_{-10.5}$ & 2.9$\pm$3.2 & 25.5$^{+2.5}_{-1.5}$\\
~~~log$M_*>$ 10.5      & 10.9$^{+0.2}_{-0.3}$ & 2.3 & 260   & 0.25$\pm$0.08 & 3.2 & 11.1$^{+0.1}_{-0.2}$ & 12.8$\pm$4.0 & 29.1$^{+1.3}_{-0.8}$\\
~~~$z>$ 2.5            &&&&&&&\\
~~~log$M_*\leq$ 10.5   & 10.3$^{+0.1}_{-0.1}$ & 2.7 & 47    &-0.05$\pm$0.19 &-0.3 & 0.0$^{+11.0}_{-0.0}$ & 0.0$\pm$10.0 & 25.5$^{+4.4}_{-2.1}$\\
~~~log$M_*>$ 10.5      & 10.9$^{+0.2}_{-0.3}$ & 2.7 & 168   & 0.11$\pm$0.10 & 1.1 & 10.8$^{+0.3}_{-0.9}$ & 6.0$\pm$5.3 & 37.3$^{+2.5}_{-1.4}$\\
\hline
IR-radio-bright QG     & 11.0$^{+0.2}_{-0.6}$ & 0.9 & 1769 & 0.26$\pm$0.03 & 8.6 & 11.1$^{+0.0}_{-0.1}$ & 11.7$\pm$1.4 & 6.4$^{+0.1}_{-0.1}$\\
~~~$z\leq$ 0.5         &&&&&&&\\
~~~log$M_*\leq$ 10.5   & 10.1$^{+0.3}_{-1.5}$ & 0.3 & 90  & 0.28$\pm$0.13 & 2.0 & 10.8$^{+0.2}_{-0.3}$ & 6.8$\pm$3.3 & 0.2$^{+0.0}_{-0.0}$\\
~~~log$M_*>$ 10.5      & 11.1$^{+0.2}_{-0.4}$ & 0.4 & 214 & 0.14$\pm$0.09 & 1.6 & 10.3$^{+0.2}_{-0.4}$ & 2.0$\pm$1.2 & 0.4$^{+0.0}_{-0.0}$\\
~~~0.5 $< z\leq$ 1.0   &&&&&&&\\
~~~log$M_*\leq$ 10.5   & 10.2$^{+0.2}_{-0.3}$ & 0.8 & 220 & 0.24$\pm$0.09 & 2.8 & 11.0$^{+0.1}_{-0.2}$ & 10.5$\pm$3.8 & 5.1$^{+0.1}_{-0.1}$\\
~~~log$M_*>$ 10.5      & 11.1$^{+0.2}_{-0.4}$ & 0.8 & 730 & 0.13$\pm$0.05 & 2.7 & 10.5$^{+0.1}_{-0.2}$ & 3.0$\pm$1.1 & 1.7$^{+0.0}_{-0.0}$\\
~~~1.0 $< z\leq$ 1.5   &&&&&&&\\
~~~log$M_*\leq$ 10.5   & 10.3$^{+0.2}_{-0.3}$ & 1.2 & 44  & 0.39$\pm$0.19 & 2.0 & 11.3$^{+0.2}_{-0.3}$ & 21.1$\pm$10.5 & 8.7$^{+1.0}_{-0.5}$\\
~~~log$M_*>$ 10.5      & 11.1$^{+0.2}_{-0.4}$ & 1.2 & 260 & 0.40$\pm$0.08 & 5.1 & 11.4$^{+0.1}_{-0.1}$ & 22.6$\pm$4.5 & 3.6$^{+0.2}_{-0.1}$\\
~~~1.5 $< z\leq$ 2.0   &&&&&&&\\
~~~log$M_*\leq$ 10.5   & 10.3$^{+0.1}_{-0.1}$ & 1.6 & 27  & 0.62$\pm$0.25 & 2.5 & 11.9$^{+0.1}_{-0.2}$ & 80.9$\pm$32.3 & 28.6$^{+4.9}_{-2.2}$\\
~~~log$M_*>$ 10.5      & 11.0$^{+0.2}_{-0.4}$ & 1.7 & 108 & 0.81$\pm$0.12 & 6.6 & 12.0$^{+0.1}_{-0.1}$ & 102.3$\pm$15.6 & 21.1$^{+0.8}_{-0.7}$\\
~~~2.0 $< z\leq$ 2.5   &&&&&&&\\
~~~log$M_*\leq$ 10.5   & 10.3$^{+0.1}_{-0.2}$ & 2.2 & 6   & 1.25$\pm$0.52 & 2.4 & 12.1$^{+0.2}_{-0.2}$ & 128.8$\pm$54.0 & 65.8$^{+14.3}_{-15.2}$\\
~~~log$M_*>$ 10.5      & 11.1$^{+0.2}_{-0.4}$ & 2.2 & 39  & 0.48$\pm$0.20 & 2.4 & 11.6$^{+0.2}_{-0.2}$ & 38.6$\pm$16.3 & 61.1$^{+5.7}_{-3.5}$\\
~~~$z>$ 2.5            &&&&&&&\\
~~~log$M_*\leq$ 10.5   & 10.5$^{+0.0}_{-0.0}$ & 2.7 & 2   & 0.42$\pm$0.90 & 0.5 & 11.4$^{+0.5}_{-11.4}$ & 28.1$\pm$60.4 & 27.8$^{+17.4}_{-9.9}$\\
~~~log$M_*>$ 10.5      & 11.0$^{+0.2}_{-0.4}$ & 2.8 & 27  & 0.44$\pm$0.25 & 1.8 & 11.5$^{+0.2}_{-0.4}$ & 30.8$\pm$17.1 & 62.5$^{+11.6}_{-6.6}$\\
\enddata
\tablenotetext{abc}{~~~The parameters follow those in Table \ref{tab:450stack}.}
\end{deluxetable*}

Overall, we see that QG candidates with 24 $\mu$m counterparts and QG candidates with 3 GHz counterparts that are not radio AGNs (i.e., IR-radio-bright QGs) exhibit the strongest stacking signal at both 450 $\mu$m and 850 $\mu$m.  These IR-radio-bright QGs account for 9.7$\pm$0.2\% (1769/18304) of all the QG candidates.  In general, we do not reach significant detections of IR-radio-faint QGs.  However, even with the low SNR, the stacked 850 $\mu$m fluxes for high-mass ($>10^{10.5}~M_{\odot}$) IR-radio-faint QGs are consistently higher than those for low-mass IR-radio-faint QGs.  This suggests that even the IR-radio-faint QGs have dust emission in the rest-frame far-IR, or they are clustered around dusty objects (see below). On the other hand, among IR-radio-bright QGs, it is not apparent that the high-mass ones show consistently higher 850 $\mu$m fluxes than the low-mass ones. This suggests that we are not seeing a population of well-behaved galaxies who follow the star-formation main sequence. This is expected for QGs.

We can compare our stacked 450 $\mu$m fluxes with the \emph{Herschel} 500 $\mu$m stacked fluxes in \citet{Man2016}.  Our mean 450 $\mu$m flux of IR-radio-faint QGs is 0.00$\pm$0.02 mJy, whose 1$\sigma$ upper limit is over 10 times lower than their stacked 500 $\mu$m fluxes of IR-faint QGs, which range from 0.2 to 2.5 mJy in different mass and redshift bins. The difference of defining the two QG subsamples (SFR derived from 24 $\mu$m to be under or over 100 $M_{\odot}$ year$^{-1}$ in \citet{Man2016}) may be one of the possible explanations. However, our mean 450 $\mu$m flux of IR-radio-bright QGs, 0.65$\pm$0.15 mJy, is still lower than most of the above mean 500 $\mu$m fluxes of IR-faint QGs (0.2 to 2.5 mJy) except for some of those with $M_*<10^{10.6}~M_{\odot}$.

Our stacked 450 $\mu$m fluxes are also lower compared with results in \citet{Magdis2021}. They selected QG candidates with several color-color diagrams and stacked samples without 24 $\mu$m detection, so we compare our stacked fluxes of IR-radio-faint QGs with their results. Their stacked \emph{Herschel} 500 $\mu$m fluxes, ranging from 0.12 to 0.59 mJy in various redshift bins, are also much higher than our stacked 450 $\mu$m flux, 0.00$\pm$0.02 mJy. On the other hand, their stacked SCUBA-2 850 $\mu$m fluxes, ranging from 0.04 to 0.1 mJy, are at a similar level as our stacked 850 $\mu$m fluxes.

A possible explanation for the differences between the SCUBA-2 450 $\mu$m and \emph{Herschel} 500 $\mu$m stacked fluxes is that the stacked \emph{Herschel} fluxes were biased by source clustering at the scale of the large $35\arcsec$ \emph{Herschel} 500 $\mu$m beam (e.g., \citealt{Viero2013}, also see discussion in \citealt{Bethermin2017}). Although \citet{Magdis2021} modeled the emission of all the stacked images to separate surrounding sources, it appears that their 500 $\mu$m stacked fluxes are higher.  Although \citet{Man2016} applied \texttt{SIMSTACK} to stack and deblend simultaneously, it remains possible that the effects of source blending and clustering were not completely removed.

The above comparison confirms the well known bias in submillimeter stacking analysis: when source are clustered at scales comparable to the beam size, the stacked flux would be overestimated. This bias becomes quite severe under \emph{Herschel}'s large beams in the two longest wavebands.  How about our SCUBA-2 stacked fluxes?  In the previous section, we showed that QG candidates are clustered around 850 $\mu$m sources at scales of SCUBA-2's beam.  So if we blindly stack these QGs in the 850 $\mu$m image, the stacked flux will be overestimated.  Fortunately, the majority of our 850 $\mu$m stacking signal comes from the IR-radio-bright subsample.  Their 24 $\mu$m and 3 GHz counterparts are likely to be 850 $\mu$m sources themselves, and the bias caused by clustering should therefore be negligible. On the other hand, the IR-radio-faint sample does not have deep high-resolution data to confirm that the QGs are responsible for the detected 850 $\mu$m fluxes.  Therefore, strictly speaking, our stacked 850 $\mu$m fluxes for IR-radio-faint QGs should be considered as upper-limits.  Even if a detection is reached on certain subsample of IR-radio-faint QGs, the detected flux should be only an indication that these QG candidates are physically related to faint submillimeter emitters, rather than evidence for in situ star formation in the QG candidates.

Finally, we can examine if the strong 850 $\mu$m detection (8.6$\sigma$) of the IR-radio-bright QGs really come from galaxies in the QG color-color space in the $NUV$--$r$--$J$ diagram, or from galaxies originally in the SFG color space scattered by photometric errors across the selection boundary. In Section \ref{sec:QG_selection}, we show that such SFG contamination caused by photometric errors account for about 7.5\% of the selected QGs.  With the same method, the estimated fraction for misidentified IR-radio-bright QGs caused by photometric errors is slightly higher, 8.9\%.  Moreover, we identified individual IR-radio-bright QGs whose probability of being scattered from the SFG color space to be $>0.05$. These sources account for 33\% of our IR-radio-bright QGs (584/1769). We excluded them and re-did the stacking on the remaining IR-radio-bright QGs, and still obtained a strong detection of $0.22\pm0.04$ mJy (5.9$\sigma$) despite the very generous probability cut of $>0.05$. These results imply that misidentified QG candidates due to photometric errors account for only $<$10\% of our estimated dusty SFG contamination, and this minor population does not dominate our stacking results. The majority of the dusty SFG contamination is caused by their intrinsic properties rather than photometric errors.

\subsection{Examining the Quiescence}

To examine if our sub-samples are consistent with a quiescent population, we need to derive their SFRs and compare with their stellar masses.

We calculated the IR luminosity from the mean submillimeter fluxes and the median redshift of each group of the stacking sample. Since we only conducted measurements at 450 and 850 $\mu$m, we performed single-band SED ``fitting'' by assuming that there is a unique relation between SED shape and IR luminosity. To do so, we adopt the luminosity-dependent dust SED templates of J.\ K.\ Chu et al.\ (in preparation), which are based on the latest \emph{WISE} and \emph{Herschel} photometry for 201 local IR-selected galaxies \citep{Chu2017}.  This set of templates covers IR luminosity of $7\times10^9$ to $1.7\times10^{12} L_{\odot}$.  We further supplement the submillimeter galaxy SED from the zLESS program \citep{Danielson2017}, which has an IR luminosity of $5.2\times10^{12} L_{\odot}$. We redshift these SEDs to the redshifts of our targets and calculated their observed 450 $\mu$m or 850 $\mu$m fluxes.  We picked the templates with redshifted fluxes closest to our stacked flux and interpolate between the template fluxes to obtain the IR luminosity of our targets. We scaled the IR luminosity by 1/SNR to estimate $1\sigma$ error of the IR luminosity. For groups with negative mean flux, we calculated the corresponding IR luminosity of flux error to estimate $1\sigma$ upper limit of the IR luminosity. The results are are presented in the seventh columns of Tables \ref{tab:450stack} and \ref{tab:850stack}.

To verify the results based on the local SEDs of Chu et al., we also repeated the calculations using the SED library of \citet{Schreiber2018c}. Overall, we find no systematic differences if we assume main sequence galaxies ($R_{\rm SB}=1$) for the Schreiber et al.\ library. The mean difference in the calculated $L_{\rm IR}$ is less than 0.1 dex for the non-zero entries in Tables \ref{tab:450stack} and \ref{tab:850stack}, while the rms dispersion is within 0.25 dex. This small difference can be further reduced if we assume a sub-main-sequence $R_{\rm SB}$ for the IR-radio-faint subsamples in Table~\ref{tab:850stack} and a starburst $R_{\rm SB}$ for the IR-radio-bright subsamples. This tuning of the $R_{\rm SB}$ parameter is consistent with our interpretation of these two subgroups (see below). In our subsequent analyses, we adopt the calculations based on the SEDs of Chu et al.

After calculating the IR luminosity, we followed the $L_{\rm IR}$--SFR calibration applied in \citet{Man2016}. We estimate the SFR by applying the relation applicable for SFGs \citep{Kennicutt1998}:
$$\mathrm{SFR} (M_{\odot}\, \mathrm{yr^{-1}})=1.7\times10^{-10} L_{\rm IR} (L_{\odot}).$$
We then adjusted the obtained SFR to the \citet{Chabrier2003} IMF by applying the calibration used in \citet{Man2016}:
$$\mathrm{SFR}_{\rm Chabrier}=\mathrm{SFR}_{\rm Salpeter}/1.7$$

The results are presented in the eighth columns of Table \ref{tab:450stack} and Table \ref{tab:850stack} as SFR$_{\rm 450 \mu m}$ and SFR$_{\rm 850 \mu m}$, respectively.  A few observations can be made here. First, the 850-$\mu$m derived SFR is in general higher than the 450-$\mu$m derived SFR. This is partly caused by the much deeper in luminosity sensitivity of the SCUBA-2 450 $\mu$m imaging and the $<3\sigma$ thresholds we imposed in the stacking procedure. If we remove this threshold in the 450 $\mu$m imaging, the difference reduces to within a factor of 2, which is not very significant if we consider the overall low S/N of the 450 $\mu$m stacked fluxes and the small number of available sources for the 450 $\mu$m stacking.

We compare our results with SFRs derived from optical SED fitting in COSMOS2015 (last column in Table \ref{tab:450stack} and  \ref{tab:850stack}). Their mean SFRs of IR-radio-faint QGs are higher than ours at high $z$ but their mean SFRs of IR-radio-bright QGs are lower.  This can be explained with the age-extinction degeneracy in the SED fitting when there is an absence of far-IR photometry.

\begin{figure*}[ht!]
\epsscale{1.15}
\plotone{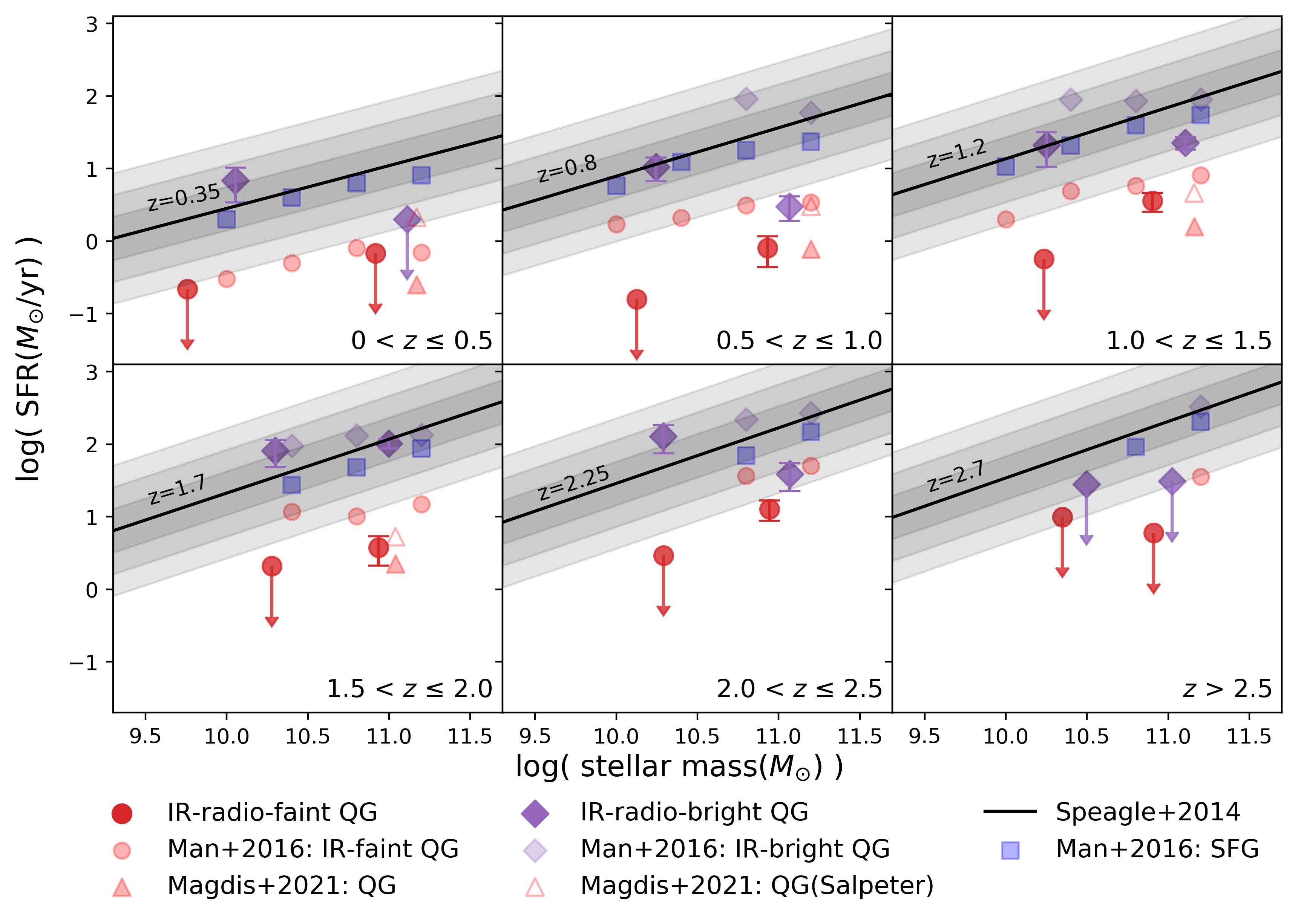}
\caption{SFR derived from 850 $\mu$m fluxes versus stellar mass. The purple diamonds represent IR-radio-bright QGs, while the red circles represent IR-radio-faint QGs. The smaller semi-transparent symbols are results from other works. The purple diamonds, red circles, and blue squares represent the IR-bright QGs, IR-faint QGs, and SFGs in \citet{Man2016}, respectively. The red triangles represent QGs in \citet{Magdis2021}. IR-radio-bright QGs in our work are defined as QG candidates either with 24 $\mu$m counterparts or with 3 GHz counterparts labeled with SFG flags in the VLA catalog \citep{Smolcic2017}, while IR-bright QGs in \citet{Man2016} are defined as QG candidates with SFR derived from 24 $\mu$m over 100 $M_{\odot}$ year$^{-1}$. QGs in \citet{Magdis2021} are defined as QG candidates without 24 $\mu$m detection. The filled triangles are derived by the same $L_{IR}$--$\mathrm{SFR}$ conversion with the other two works, while the open triangles are derived by the conversion described in their work.} The black solid line shows the SFR of the redshift-dependent main sequence with a 1--3 $\times$ 0.3 dex scatters in \citet{Speagle2014}.\label{fig:SSFR}
\end{figure*}

We compare the submilimeter-derived SFRs with the stellar masses of the galaxies in Fig.~\ref{fig:SSFR}. We show the star-formation ``main sequence'' of \citet{Speagle2014} with black solid lines and the $\pm$0.9 dex ranges with shaded areas.  

We show the results from \citet{Man2016} for comparison (Fig.~\ref{fig:SSFR}). Our results are in broad agreement with theirs but tend to have slightly lower SFR for low-$z$ samples. In the $0.5<z\leq1.0$ and $1.0<z\leq1.5$ redshift bins, their massive IR-bright QGs are located on or $\sim1$ to $2\sigma$ above the main sequence, while our IR-radio-bright QGs are located $\sim2$ to $3\sigma$ below the main sequence. Their IR-faint QGs are located $\sim3\sigma$ below the main sequence, while our IR-radio-faint QGs are located $>3\sigma$ below the main sequence.  Other than these, our derived SFRs are fairly consistent. We note that the SFR of \citet{Man2016} was derived from SED fitting using stacked fluxes across the entire far-IR range.  This explains why their 500 $\mu$m stacked fluxes are much higher than ours, but their SFRs are not. 

We also show the results from \citet{Magdis2021} in Fig.~\ref{fig:SSFR}. Their SFRs are about the same as or slightly lower than our SFRs, different from the comparison with \citet{Man2016}. We note that their IR luminosities were derived from SED fitting using stacked fluxes from mid-IR to radio data. They then also obtained SFRs by applying the relation in \citet{Kennicutt1998}, but they used a Salpeter IMF and added SFR derived from the optical photometry. We converted their IR luminosity to SFR by the same process in \citet{Man2016} and this work for a fair comparison.  We also show their SFRs obtained by the original conversion in their work for reference, which are in general closer to the SFRs from \citet{Man2016}.

The conclusion we can draw from Fig.~\ref{fig:SSFR} is that the IR-radio-faint QGs are in general below the star-formation main sequence, while the majority of the IR-radio-bright QGs are consistent with the main sequence, probably except for the high-mass end in the two low-redshift bins and in the highest redshift bin.

To sum up, our stacking results show that only the IR-radio-bright QGs have SFR similar to main-sequence galaxies. These are likely to be faint dusty SFGs that contaminate the QG color selection. However, the population of the IR-radio-bright QGs is small, which accounts for 9.7$\pm$0.7\% (179/1846) and 9.7$\pm$0.2\% (1769/18304) of all the  QG candidates, respectively, in the 450 and 850 $\mu$m images. The fractions range from 7\% to 12\% in different redshift bins and do not have a strong redshift dependence. We conclude that the contamination of dusty SFGs is of $\sim$ 10\% among the color-selected QG candidates, and that the contamination can be removed using multi-wavelength data such as the 24 $\mu$m and 3 GHz data for the COSMOS field.

For comparison, \citet{Man2016} suggested that the maximum contamination is 15\% and could be removed by using 24 $\mu$m observations. In this study, we used submillimeter data with better sensitivities and resolutions, and our estimate of contamination is somewhat tighter (10\%) than that in \citet{Man2016}. 

Like what we did in Section~\ref{subsubsec:traditional_matching_process}, if we assume the same fraction of QGs among AS2COSMOS sources, we can estimate the number of faint submillimeter sources that are QGs.  For the number of faint submillimeter sources, we again applied the 850 $\mu$m number count in \citet{Simpson2019} and extrapolated it to a flux level of S$_{850 \mu m}= 0.5$ mJy.  This leads to 707$\pm$462 QGs among faint submillimeter sources, and a dusty galaxy contamination rate among QGs of 3.9$\pm$2.5\%.  We can further extrapolate the counts to 0.26 mJy, the stacked 850 $\mu$m flux of IR-radio-bright QGs. This will further increase the estimated contamination rate. However, such an extrapolation is is probably too aggressive given the uncertainty in the faint-end counts. Nevertheless, considering the unknown uncertainty of extrapolating the number count to a flux level lower than the detection limit, we concluded that the above estimation is not inconsistent with the $\sim10\%$ contamination derived from the stacking of IR-radio-bright QGs.

Finally, using either 24 $\mu$m or 3 GHz data to pinpoint star-forming contaminants among color-selected QG candidates may not work well at high redshifts ($z>3$ or 4). This is because mid-IR and radio suffer from the strong $K$-correction and are not sensitive to high-redshift SFGs.  Our $3.4\sigma$ detection in the 850 $\mu$m stacking of the IR-radio-faint QGs at $z>2$ seems to agree with this, i.e., there may exist dusty galaxy contamination that are faint in the mid-IR and radio.  In other words, the effectiveness of QG color selection at high redshift remains untested in this framework.  Since the formation of QGs at higher redshifts require both rapid growth of the stellar population and rapid quenching, identifications of high-redshift QGs are of great interest \citep{Merlin2018,Straatman2014,Carnall2020, Valentino2020}.  Removing dusty contaminants among high-redshift QGs is beyond the sensitivities of \emph{Spitzer}, \emph{Herschel}, and the current VLA, and will require deep ALMA data.

\section{AGN Properties} \label{sec:AGN_properites}

\begin{figure*}[!ht]
\epsscale{1.15}
\plotone{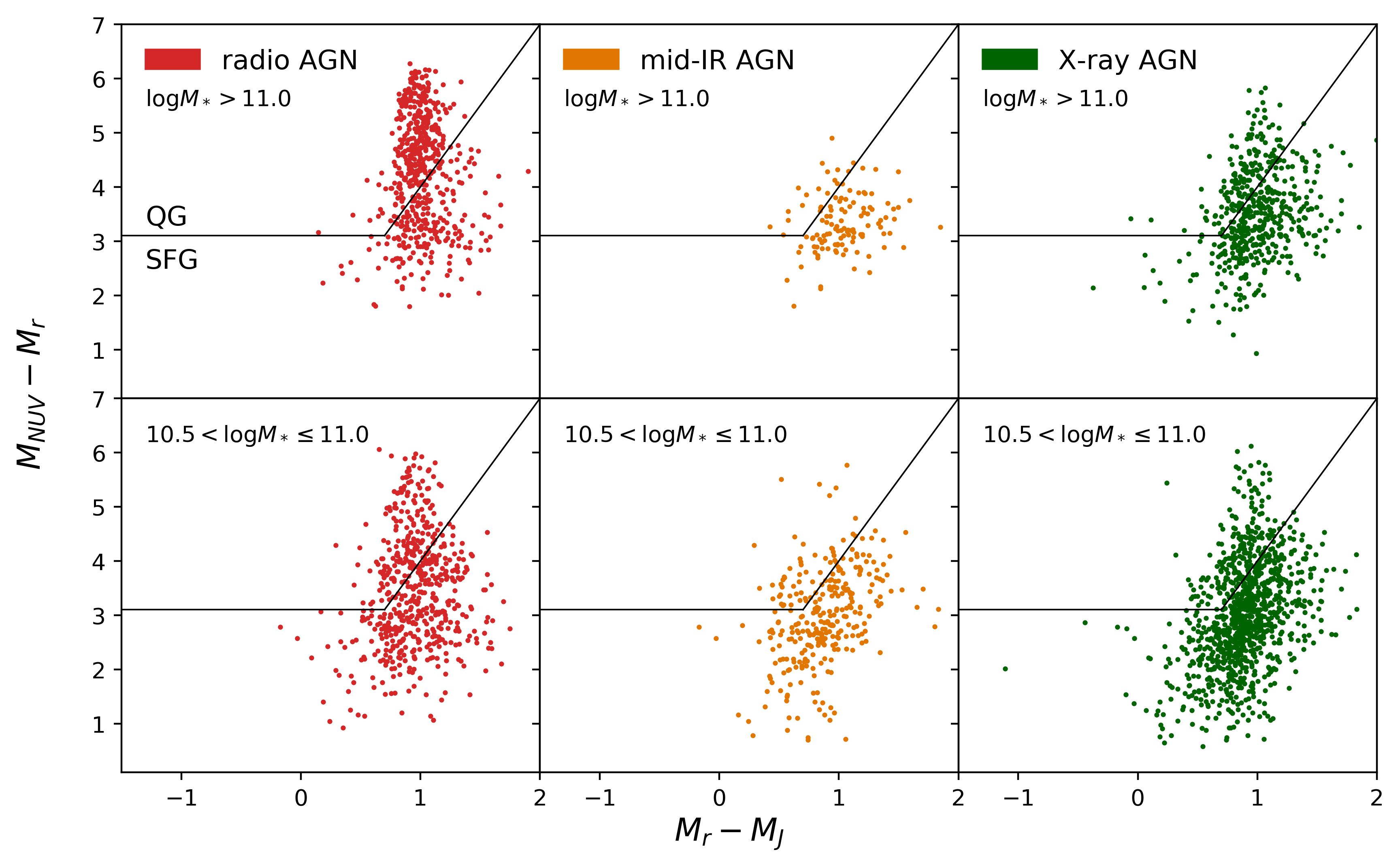}
\caption{The distribution of radio AGNs (left penals), mid-IR AGNs (middle penals), and  X-ray AGNs (right penals) on the $NUV$--$r$--$J$ diagram with $M_*>10^{11}~M_{\odot}$ (top penals) and $10^{10.5}~M_{\odot}<M_*\leq10^{11.0}~M_{\odot}$ (lower penals).\label{fig:NUVrJ_AGN}}
\end{figure*}

In this section, we discuss the properties of the AGNs among our QG candidates. Fig.~\ref{fig:NUVrJ_AGN} shows the distribution of three different classes of AGN samples in the $NUV$--$r$--$J$ diagram, including radio AGNs, mid-IR AGNs, and X-ray AGNs (Section~\ref{subsec:AGN_samples}) in two mass bins. The stellar mass of the samples were limited to above $10^{10.5}~M_{\odot}$. This is because the samples are likely to be incomplete below $10^{10.5}~M_{\odot}$ at $z\sim2$ (see Fig.~\ref{fig:data} (b)).  This mass limit is also consistent with the 90\% completeness limit found by \citet{Laigle2016} for QGs at high redshifts. We therefore applied this stellar mass cut for fair comparisons of the QG fractions. In Fig.~\ref{fig:NUVrJ_AGN}, we can see that the distribution of the radio AGNs is different from those of the other two. A similar distinction also exists between radio selected sources and 24 $\mu$m selected sources in Fig.~\ref{fig:NUVrJ243}. Fig.~\ref{fig:NUVrJ_AGN} provides evidence that the difference in Fig.~\ref{fig:NUVrJ243} is driven by radio AGNs.

\begin{figure}[ht!]
\epsscale{1.15}
\plotone{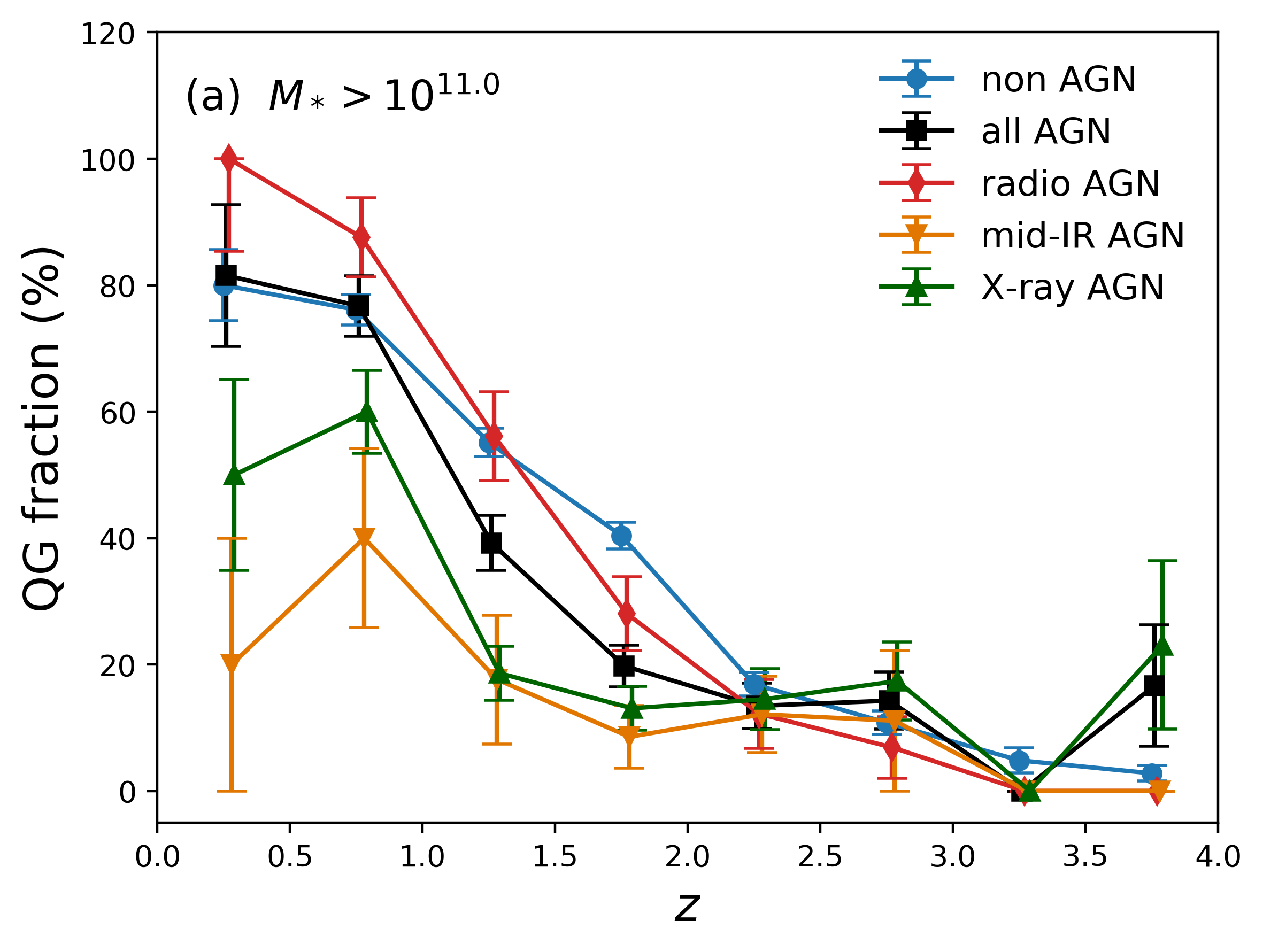}
\epsscale{1.15}
\plotone{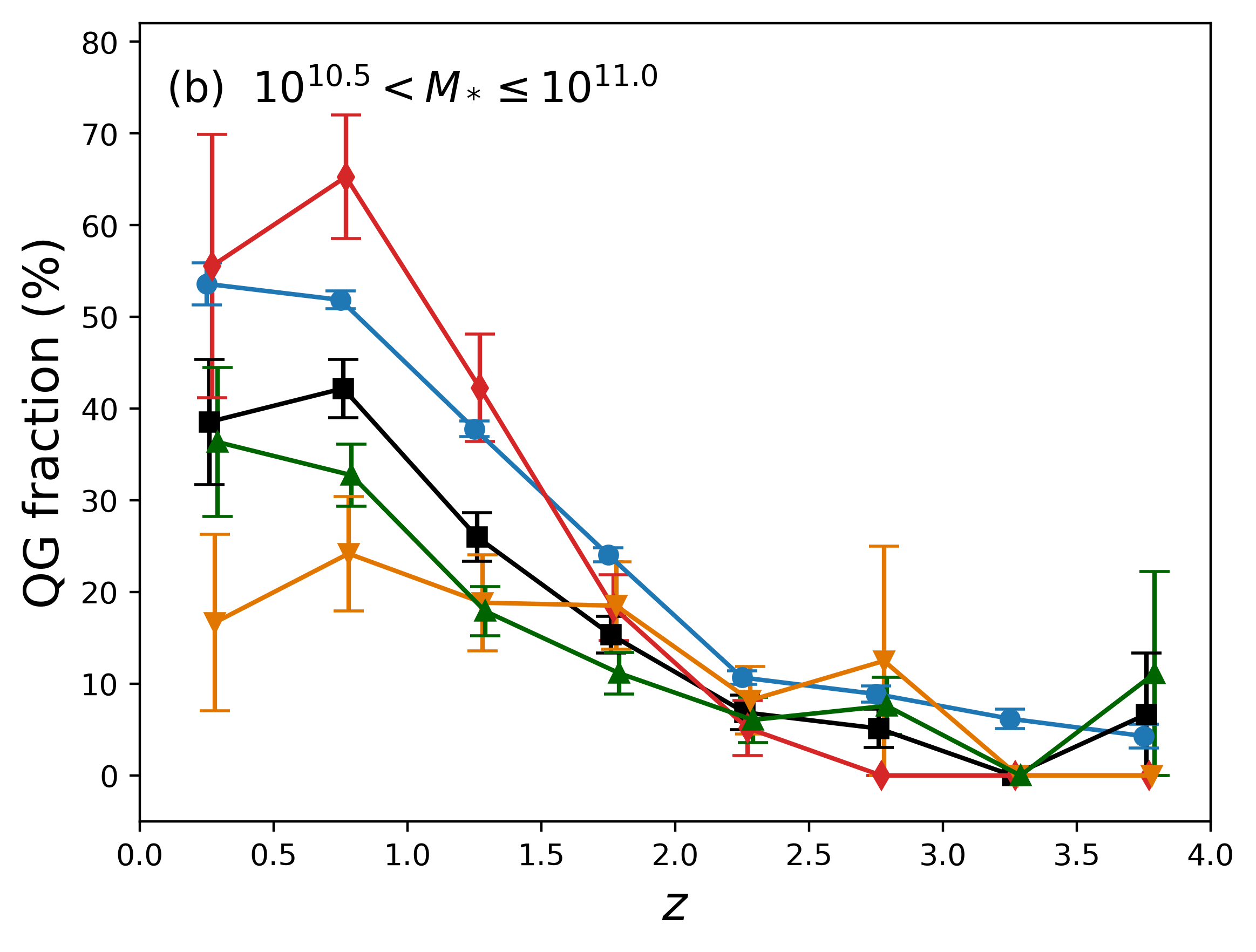}
\caption{The QG fraction among different classes of AGN samples in redshift bins with $M_*>10^{11}~M_{\odot}$ (a) and $10^{10.5}~M_{\odot}<M_*\leq10^{11.0}~M_{\odot}$ (b). The data points are slightly offset along x-axis for clarity. The errors are Poissonian.\label{fig:QG_fraction}}
\end{figure}

We calculated the QG fractions in different classes of the AGN samples in different redshift and mass bins (Fig~ \ref{fig:QG_fraction}). The QG fraction of the radio AGNs is ($\sim0.1$ to $1.9\sigma$) higher than that of the non-AGN samples at $z<1.5$ in both two mass bins, while the QG fractions of the other two AGN classes are significantly lower. The trend does not persist beyond $z\sim1.5$, and this may be due partially to the detection limit of the radio AGNs, or a real redshift evolution. The high QG fraction among radio AGNs implies a correlation between radio jets and our QG candidates.

\begin{figure}[ht!]
\epsscale{1.15}
\plotone{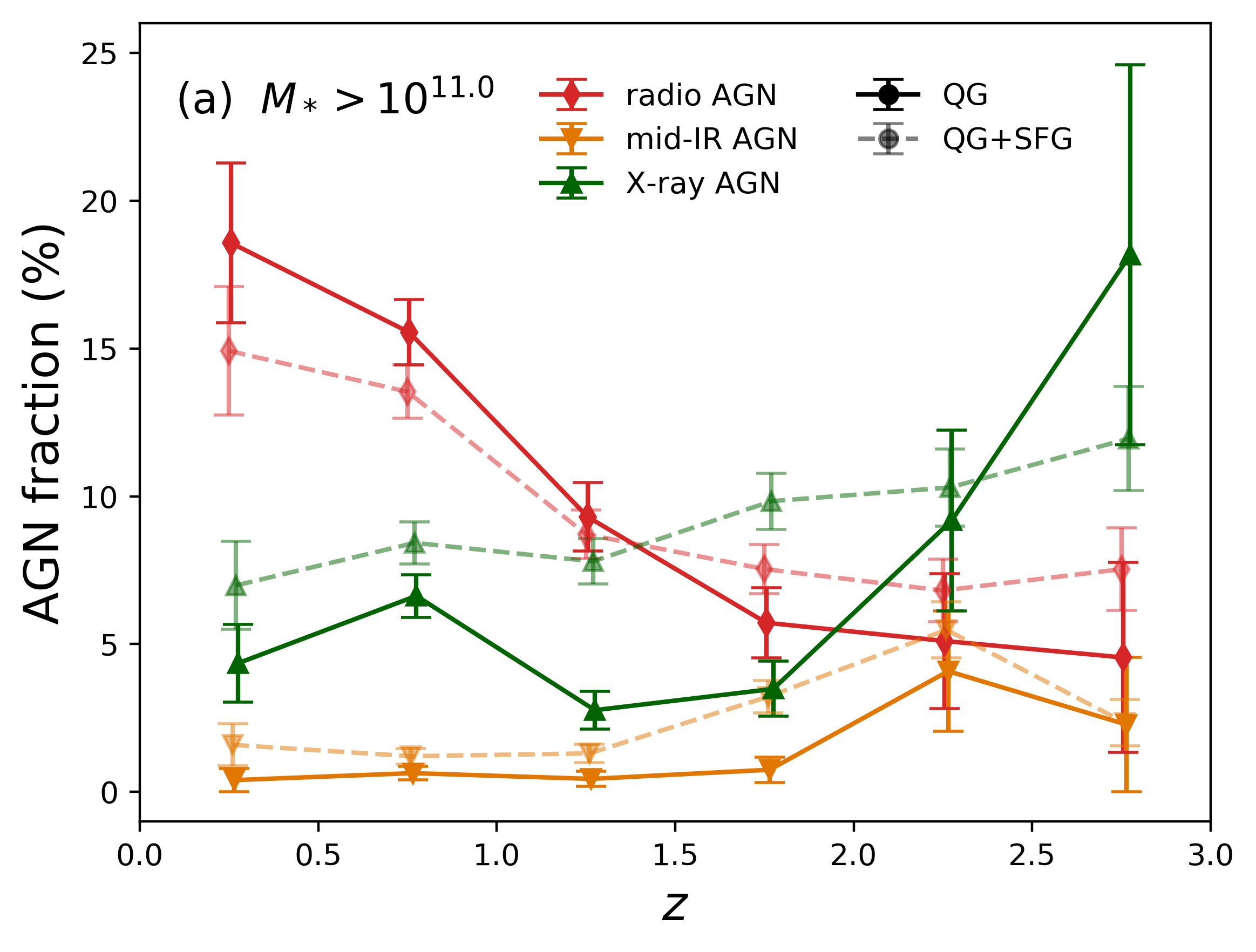}
\epsscale{1.15}
\plotone{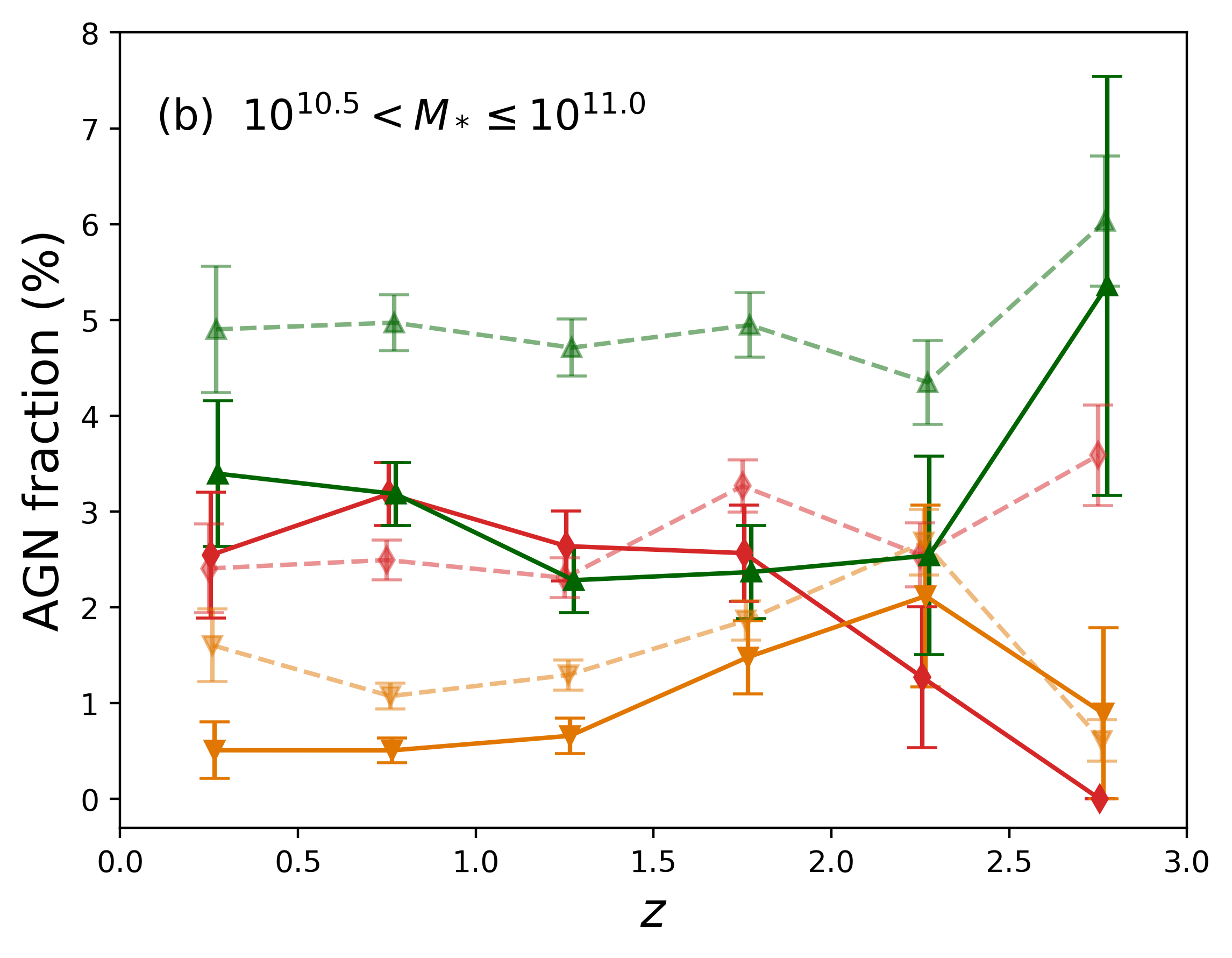}
\caption{Different classes of AGN fraction among COSMOS2015 optical sample in redshift bins with $M_*>10^{11}~M_{\odot}$ (a) and $10^{10.5}~M_{\odot}<M_*\leq10^{11.0}~M_{\odot}$ (b). The solid lines represent AGN fraction among QG candidates, while the dotted line represent AGN fraction among all the galaxies (QGs and SFGs). The data points are slightly offset along x-axis for clarity. The errors are Poissonian.\label{fig:AGN_fraction}}
\end{figure}

We also calculated the AGN fractions in our COSMOS2015 sample (Fig~ \ref{fig:AGN_fraction}). The radio AGN fraction among QG candidates is ($\sim0.2$ to $1.7\sigma$) higher than that among the COSMOS2015 QG and SFG sample, also at $z<1.5$ and in both two mass bins. The situations are the opposite for both the mid-IR and X-ray AGNs, where the AGN fractions among the QG+SFG sample are higher. The higher radio AGN fraction among QGs also suggests a correlation between radio jets and our QG candidates.

The above correlation between radio AGNs and QGs may provide some clues about the quenching mechanism. Our data suggest that the formation of radio jets and quenching of star formation are connected, but the exact causality is unclear. One possible scenario is that radio-mode AGN feedback, particularly radio jets with high kinetic energy, could have impacts on the host galaxy and its environment \citep{Croton2006, Bower2006, Fabian2012, Somerville2008, Wagner2011}. Either the interstellar medium in the host galaxy may be disrupted by the jets \citep[e.g.,][]{Fabian2012,Somerville2008}, or the gas in the halo can be heated up and be prevented from cooling and falling onto the host galaxy \citep[e.g.,][]{Croton2006, Bower2006, Somerville2008}. Although the radio AGNs fraction is only $\sim$ 10 to 20\% among QG candidates (Fig.~\ref{fig:AGN_fraction}), this can be explained by the AGN duty cycle in the massive galaxies.

On the other hand, it is also possible that radio AGNs are not directly linked to the quenching process.  Instead, it may be just easier to trigger radio AGNs in galaxies that are already quenched.  In the local universe, it is well known that there is a dominant tendency for radio galaxies to reside in early-type hosts \citep[e.g.,][]{Govoni2000,Hamilton2002}, which is consistent with what we found on QGs and radio AGNs.  The lack of cool gas in early-type (quenched) galaxies can lead to radiatively inefficient accretion that is associated with the radio mode AGN feedback \citep{Croton2006, Bower2006, Somerville2008}.

Observationally, the stacking result in \citet{Man2016} also showed a similar trend. The authors compared the radio stacked fluxes and the \emph{Herschel} stacked fluxes of their sample. Their SFRs derived from the two are consistent for SFGs, but there exists a radio excess for massive IR-faint QGs with $M_*\geq10^{11}~M_{\odot}$ at $z<$ 1.5. They further measured the FIR-to-radio luminosity ratio and concluded that most of their massive QGs host low-luminosity radio AGNs. Also, \citet{Smolcic2009} found that the radio AGN fraction among red galaxies increases from $z \sim$ 0.5 to 1, showing that radio AGNs play an important role at $z \sim$ 1. It is also known that SFR is correlated with AGN accretion rate \citep{Zhuang2020}, and that low-accretion rate AGNs are radio-loud \citep{Ho2002, Ho2008}. This implies that QGs tends to be radio AGNs. All these results suggest that there is a correlation between QGs and synchrotron radiation from radio AGNs.

We could also see a weak correlation between X-ray AGNs and QG candidates at $z>2.5$. In Fig.~\ref{fig:QG_fraction}, the QG fractions among X-ray AGNs are $\sim0.6$ to $1.5\sigma$ larger than those among non AGNs in $z>2.5$ redshift bins, with respect to their own error bars. In Fig.~\ref{fig:AGN_fraction}, the AGN fraction among QGs is $\sim0.9\sigma$ larger than that among the full sample in the $z>2.5$ and $M_*>10^{11}~M_{\odot}$ bin. This may be caused by either selection bias or a real evolution trend. The evolution trend could be explained by the role of quasar-mode AGN feedback \citep{Fabian2012, Somerville2008}, gas inflow in X-ray AGNs that removes gas and quenches star formation. One possibility is that the mode of AGN quenching may change from quasar-mode to radio-mode from high $z$ to low $z$. Another possibility could be that X-ray AGNs are related to the initial quenching, while radio AGNs are responsible for the maintenance of the quiescence. This could also explain the increasing radio AGN fraction among QG candidates in Fig.~\ref{fig:AGN_fraction} at lower redshift. Nevertheless, the rises in the X-ray AGNs in Fig.~\ref{fig:QG_fraction} and \ref{fig:AGN_fraction} only occur in the highest redshift bins where the sample sizes are the smallest and the selection completeness is less well understood. This has to be further tested with more data and careful examination of various selection biases in the high-redshift ends.

To sum up, our data show a strong correlation between radio AGNs and QGs but do not point to the right scenario. Our data also do not show whether radio AGNs are related to the initial quenching, or just related to the maintenance of the quiescence.

\section{Summary} \label{sec:summary}

In this study, we examined the submillimeter properties of $NUV$--$r$--$J$ selected QG candidates at $z\lesssim3$. We cross-matched the QG candidates with bright submillimeter sources detected by JCMT SCUBA-2 and ALMA. For the former, we used \emph{Spitzer} 24 $\mu$m and VLA 3 GHz data to refine their positions to overcome the low angular resolution of JCMT.  This way, we found that 0.16$\pm$0.03\% to 0.43$\pm$0.15\% among our QG candidates are likely to be bright 850 and 450 $\mu$m submillimeter galaxies, respectively. The contamination increases to 1.72$\pm$0.50\% to 3.51$\pm$2.48\% at $z>$ 2.  We further performed stacking analysis of QG candidates in the JCMT 450 and 850 $\mu$m images.  We can obtain strong stacking detections on a subsample of QGs with \emph{Spitzer} 24 $\mu$m and VLA 3 GHz counterparts that are not radio AGNs.  This special class of ``IR-radio-bright'' QGs account for about 10\% of the entire QG sample and they are likely to be faint submillimeter sources with SFRs of a few tens to about a hundred $M_\sun$ yr$^{-1}$.  These results are broadly consistent with the contaminate rates derived from a small sample of ALMA detected QGs and the 850 $\mu$m number counts.  We conclude that the dusty star-forming galaxy contamination rate among $NUV$--$r$--$J$ selected QG candidates is up to $\sim10\%$, but such contamination can be removed by 24 $\mu$m, submillimeter, or 3 GHz observations at current sensitivity levels.

When we cross-matched the QG candidates with JCMT SCBUA-2 850 $\mu$m SMGs without relying on high-resolution data, we adopted a large matching radius of $7\arcsec$ because of the large SCUBA-2 beam size.  This leads to a large fraction of chance projections among the matched QGs. We estimated the number of chance projections with simulations by assuming random spatial distributions for SCUBA-2 sources. After statistically subtracting the chance projections, we found that on average, 0.096 (35.4/370) QG is physically related to an 850 $\mu$m selected SMG, while ALMA observations indicate that only 0.026 (9.7/370) QG really coincides with an SMG within $1\arcsec$. This implies a clustering between these two populations at a scale of $1\arcsec$ to $7\arcsec$, and should be a future topic of investigation.

Finally, we examined the QG fractions among our AGN samples and found a correlation between our QG candidates and radio AGNs. When we limited our studies to galaxies with stellar masses larger than $10^{10.5}M_\sun$, we found that the QG fraction of radio AGNs are larger than those of the non-AGN samples, IR AGNs, and X-ray AGNs at $z<$ 1.5. This suggests a connection between the radio jets and the quenching or the maintenance of the quiescence of the QGs, or the so-called radio-mode AGN feedback. However, our data do not rule out the possibility that radio AGNs are just more easily triggered in quenched galaxies, rather than being responsible for the initial quenching.

\acknowledgments

The authors thank Bau-Ching Hsieh, Ian Smail, Iary Davidzon, and Olivier Ilbert for the discussion and comments, the anonymous referee for the comments that greatly improve the manuscript, and JCMT staff for the observational support. Y.H.H., W.H.W., Y.Y.C., C.F.L., and Z.K.G. acknowledge grant support from the Ministry of Science and Technology of Taiwan (MoST, 105-2112-M-001-029-MY3, 108-2112-M-001-014-, and 109-2112-M-001-011-).  C.C.C. acknowledges MoST grant 109-2112-M-001-016-MY3. M.J.M. acknowledges the support of the National Science Centre, Poland through the SONATA BIS grant 2018/30/E/ST9/00208. M.P.K. acknowledges support from the First TEAM grant of the Foundation for Polish Science No. POIR.04.04.00-00-5D21/18-00. L.C.H. was supported by the National Science Foundation of China (11721303, 11991052) and the National Key R\&D Program of China (2016YFA0400702). Y.G. acknowledges National Science Foundation of China (NSFC) grants \#11861131007, 12033004, and 11420101002, and Chinese Academy of Sciences Key Research Program of Frontier Sciences (Grant No. QYZDJ-SSW-SLH008). The submillimeter data used in this work include observations from the JCMT Large and Legacy Programs: S2COSMOS (M16AL002), STUDIES (M16AL006), and S2CLS (MJLSC01), the JCMT PI program of Casey et al.\ (M11BH11A, M12AH11A, and M12BH21A), the ALMA program AS2COSMOS (ADS/JAO.ALMA \#2016.1.00463.S), and various ALMA archival data.  The James Clerk Maxwell Telescope is operated by the East Asian Observatory on behalf of the National Astronomical Observatory of Japan; the Academia Sinica Institute of Astronomy and Astrophysics; the Korea Astronomy and Space Science Institute; and the Operation, Maintenance and Upgrading Fund for Astronomical Telescopes and Facility Instruments, budgeted from the Ministry of Finance (MOF) of China and administrated by the Chinese Academy of Sciences (CAS), as well as the National Key R\&D Program of China (No. 2017YFA0402700). Additional funding support is provided by the Science and Technology Facilities Council of the United Kingdom and participating universities in the United Kingdom and Canada. ALMA is a partnership of ESO (representing its member states), NSF (USA), and NINS (Japan), together with NRC (Canada), NSC and ASIAA (Taiwan), and KASI (Republic of Korea), in cooperation with the Republic of Chile. The Joint ALMA Observatory is operated by ESO, AUI/NRAO, and NAOJ.

\appendix

\section{Lensed System} \label{appendix:lensed}

\begin{figure}[ht!]
\epsscale{0.5}
\plotone{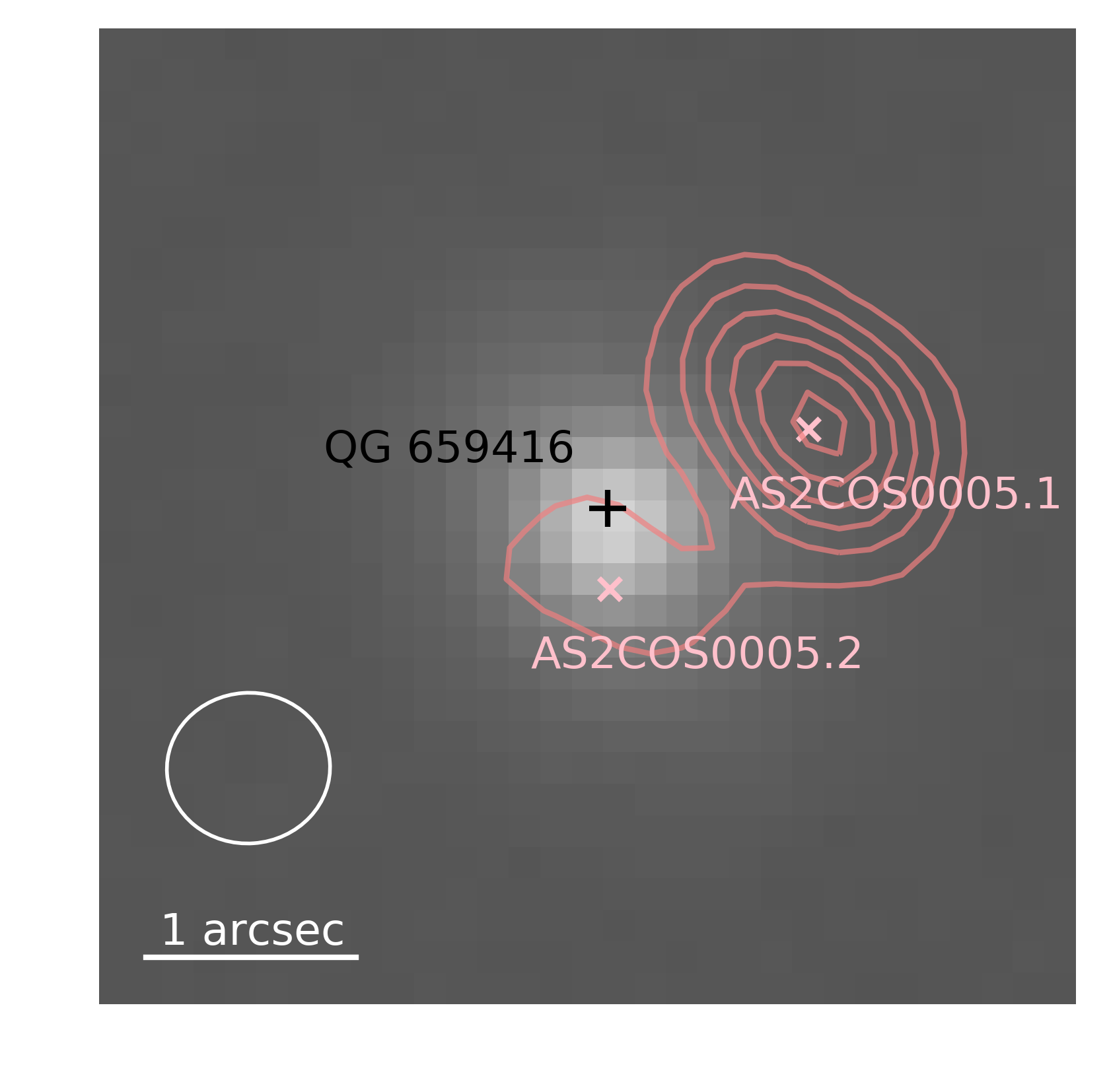}
\caption{\label{fig:lensing}CFHT $K_S$ band image and ALMA 343 GHz emission contours (5,10,...$\times \sigma$) of the lensed system. The image has a size of $\sim 4.5\arcsec$ on a side. The white circle shows the ALMA beam. The black plus symbol shows the position of the QG candidate. The pink cross symbols show the positions of the ALMA sources, which are likely to be background galaxies lensed by the QG candidate or even multiple lensed images of a single background galaxy.}
\end{figure}

One of the 850 $\mu$m detected QG candidates (ID = 659416 in the COSMOS2015 catalog; star symbol in Fig.~\ref{fig:NUVrJ_submm}) is likely to be a lensed system because of its unusual submillimeter/radio flux ratio. The $K_S$-band image and the ALMA image of the lensed system are shown in Fig.~\ref{fig:lensing}. The QG candidate is at $z = 0.36$ and is surrounded by two ALMA Band-7 sources, AS2COS0005.1 and AS2COS0005.2. The two ALMA sources have 343 GHz fluxes of 8.84 mJy and 2.20 mJy, respectively. The 3 GHz fluxes are 27.0 $\mu$Jy and 7.9 $\mu$Jy. These values lead to very similar 343-to-3 GHz flux ratios of 0.33, and 0.28. Under the framework of ``millimetric redshift'' \citep{Carilli1999}, the similar flux ratios imply similar redshifts. The estimated redshifts are 2.7 and 2.6, if we assume a radio spectral slope of 0.8 to infer the 1.4 GHz fluxes and if we adopt the millimetric redshift formula of \citep{Barger2000}:
$$z+1=0.98\times(S_{850\mu m}/S_{1.4GHz})^{0.26}.$$
The estimated redshifts are much higher than the QG photometric redshift. Therefore, the two ALMA sources are likely to be background sources, rather than dust emission from the QG candidate. Indeed, the elongated morphology of AS2COS0005.1 (Fig.~\ref{fig:lensing}) and the relative positions of the two ALMA sources strongly suggest that they are multiple images of the same background dusty galaxy lensed by the foreground QG.

This target was also identified as a lensed system in \citet{Bertoldi2007} by using the same submillimeter/radio flux ratio approach. The authors measured the Max-Planck Millimeter Bolometer Array (MAMBO-2) 250 GHz flux of the target (labeled as MM J100024+021748 or Cosbo-7) and obtained  $z\sim2.8$, consistent with our estimation. \citet{Jin2018} also reported this lensed system (labeled as ID20003080) and derived a photometric redshift of $4.0\pm0.6$ from 24 $\mu$m to radio SED fitting. In our study, we further confirmed their results by including high-resolution ALMA image, which resolved the background target into two sources with elongated morphology.

We note that there are in total three QG candidates with photometric redshift lower than 1 among those matched to SCUBA-2 sources through ALMA catalogs, and we examined their millimetric redshifts. Besides the target discussed above, the other two low $z$ targets also show higher submillimeter/radio flux ratio. However, we do not further discuss the two targets since the estimated millimetric redshifts are not significantly high.

\bibliographystyle{yahapj}
\bibliography{references}


\end{document}